\documentclass[journal,twocolumn,10pt]{IEEEtran}
\usepackage{amssymb}
\usepackage{cite}
\usepackage{graphicx}
\usepackage{psfrag}
\usepackage{subfigure}
\usepackage{url}
\usepackage{stfloats}
\usepackage{amsmath}
\usepackage{float}
\usepackage{bm}

\usepackage{mathrsfs}
\usepackage{subeqnarray}
\usepackage{cases}
\usepackage{stfloats}
\usepackage{subeqnarray}
\usepackage{cases}

\usepackage{algorithm}
\usepackage{algorithmic}

\usepackage{xcolor}
\usepackage{soul}

\usepackage[T1]{fontenc}
\usepackage[font=small,labelfont=bf,tableposition=top]{caption}
\usepackage{booktabs}
\usepackage{siunitx}

\definecolor{txcolor}{rgb}{0.8,0.09,0.3}

\ifCLASSINFOpdf
\else
\fi

\allowdisplaybreaks[3]

\begin{document}
\title{Wireless Information and Energy Transfer for Two-Hop Non-Regenerative MIMO-OFDM Relay Networks}

\author{Ke~Xiong,~\IEEEmembership{Member,~IEEE}, ~Pingyi~Fan,~\IEEEmembership{Senior Member,~IEEE},~Chuang Zhang,\\and ~Khaled Ben Letaief,~\IEEEmembership{Fellow,~IEEE}\\

\thanks{This work was supported by  the National Nature Science Foundation of China, no. 61201203,
  partly by ``973'' program, no. 2012CB316100(2), and also by the Open Research Fund of National Mobile Communications Research Laboratory, Southeast University (no. 2014D03).}

\thanks{K.~Xiong is with the School of Computer and Information Technology, Beijing Jiaotong University, Beijing 100044, China and was with the Department
of Electrical Engineering, Tsinghua University, Beijing 100084, R.P. China, e-mail:  kxiong@bjtu.edu.cn, kxiong@tsinghua.edu.cn.

P. Y. Fan, and C. Zhang are with the Department of Electrical Engineering, Tsinghua University, Beijing, R.P. China, 10008, e-mail:  fpy@tsinghua.edu.cn,\{Dongyq08@, Taoli\}@mails.tsinghua.edu.cn.

K. B. Letaief is with the School of Engineering, Hong Kong University of Science \& Technology (HKUST). e-mail:  eekhaled@ece.ust.hk.
}}

\maketitle

\begin{abstract}
This paper investigates the simultaneous wireless information and energy transfer for the non-regenerative multiple-input multiple-output orthogonal frequency-division multiplexing (MIMO-OFDM) relaying system. By considering two practical receiver architectures, we present two protocols, time switching-based relaying (TSR) and power splitting-based relaying (PSR). To explore the system performance limit, we formulate two optimization problems to maximize the end-to-end achievable information rate with the full channel state information (CSI) assumption. Since both problems are non-convex and have no known solution method, we firstly derive some explicit results by theoretical analysis and then design effective algorithms for them. Numerical results show that the performances of both protocols are greatly affected by the relay position. Specifically, PSR and TSR show very different behaviors to the variation of relay position. The achievable information rate of PSR monotonically decreases when the relay moves from the source towards the destination, but for TSR, the performance is relatively worse when the relay is placed in the middle of the source and the destination. This is the first time to observe such a phenomenon. In addition, it is also shown that PSR always outperforms TSR in such a MIMO-OFDM relaying system. Moreover, the effect of the number of antennas and the number of subcarriers are also discussed.
\end{abstract}

% \fi
\IEEEpeerreviewmaketitle

\section{Introduction}

Energy harvesting (EH) is capable of powering communication devices
and networks with energy harvested from environment, which has emerged as a
promising approach to prolong the lifetime of energy constrained
wireless communication\cite{R:X1,R:N1,R:N2}. For instance, in wireless sensor networks, when a sensor is depleted of energy, it cannot
 fulfill its role any longer unless the source of energy is replenished.
Although replacing or recharging batteries provides a solution to this problem, it may incur a
high cost and sometimes even be unavailable due to some physical or economic limitations (e.g., in a toxic environments and
for sensors  inside the body or embedded in building structures)\cite{R:X2}. Comparatively, harvesting energy from external environment may provide a much safer and much more convenient solution for such kinds of scenarios.

\subsection{Background}
As for energy harvesting techniques, the primary ones\cite{R:X3,R:E1,R:E2,R:E3} rely on external energy sources, such as solar, wind, vibration, thermoelectric effects or other physical phenomena. Since these energy sources are not components of the communication network, to use primary EH techniques requires the deployment of peripheral equipments to harvest external energy. Moreover, the external energy source in most cases cannot be controlled and thus not be always available. Such uncertainty is too critical for the practical scenario that has high requirements on reliability and stability, so it limits the applications of conventional EH techniques. Recently, a new branch of EH techniques has been presented, in which the receiver is able to collect energy from ambient radio frequency (RF) signals and the wireless signal is used
as a media to deliver information and energy simultaneously \cite{R:X4,R:X5,R:WP1,R:WP3,R:X6,R:EP1}. Thus, it potentially provides great convenience to mobile users \cite{R:EP1}.

\subsection{Previous work}
As a matter of fact, the
concept of simultaneous wireless information and energy transfer (SWIET) can be traced back to \cite{R:X4}, where the tradeoff between the energy and information rate was also characterized for the
point-to-point communication scenario. The extension of SWIET to frequency selective channels was studied in \cite{R:X5}. Later, some works, see e.g., \cite{R:WP1,R:WP3}, investigated the SWIET for other scenarios including multi-antenna systems \cite{R:WP1} and multi-user systems\cite{R:WP3}. In these works, the EH receiver was assumed to be able to simultaneously observe and extract power from the same received signal.

However, the authors in \cite{R:EP1} pointed out that this assumption may be not well available in practice, because practical circuits for harvesting energy from RF signals are not yet able to decode the carried information directly (power collection and information receiving operating with very different power sensitivity, e.g., -10dBm for energy receivers versus -60dBm for information receivers).
Therefore, they proposed an implementable design with separated information decoding and energy harvesting receiver for SWIET in \cite{R:EP1}, where two practical receiver architectures, namely, \textit{time switching} (TS) and \textit{power splitting} (PS) were presented. With TS employed at the receiver, the received signal is either processed by an energy receiver for energy harvesting
or processed by an information receiver for information decoding (ID). With PS employed at the receiver, the received signal is split into two signal flows with a fixed power ratio
by a power splitter, where one stream is to the energy receiver and
the other one is to the information receiver.

Due to their implementable features, TS and PS architectures for SWIET recently have attracted much attention, see e.g., \cite{R:WP4,R:WP5}. In \cite{R:WP4}, it investigated
the joint optimization of transmit power control and scheduling
for information and energy transfer with the receiver's mode
switching over flat fading channels. In \cite{R:WP5}, the authors focused on exploring the problem of throughput
optimization for the save-then-transmit protocol with variable energy harvesting rate.

Since in wireless cooperative or sensor networks, the relay or sensor nodes often have very limited battery storage and require some external charging mechanisms to remain active in the network. Energy
harvesting for such networks seems particularly important as it can enable information relaying for the transmissions. Thus, some recent works began to stress the SWIET with separated energy harvesting and information receiving architecture for relay systems, see e.g., \cite{R:EP2,R:EP3,R:D2,R:HC}. In \cite{R:EP2}, EH policies were designed for one-way relaying system, where non-regenerative relaying protocol was involved and the outage probability, as well as the ergodic capacity, was analyzed. In \cite{R:EP3}, power allocation strategies were investigated for multiple source-destination pair cooperative EH relay networks, where non-regenerative protocol was also considered. In \cite{R:D2}, the SWIET was considered for cooperative networks with spatially random relays, where the outage and diversity performance were studied by applying stochastic geometry, and in \cite{R:HC}, the distributed PS-based SWIET was designed for interference  relay channels by using game theory. However, these works just investigated the SWIET in single-carrier and single antenna relay systems.

It is well known that multiple-input multiple-output (MIMO) and orthogonal
frequency-division multiplexing (OFDM) have emerged
as two effective solutions to achieve high spectral
efficiency and throughput for future broadband wireless systems\cite{R:MIMO,R:OFDMf,R:OFDMF2}. Therefore, some recent works also began to discuss the
SWIET for MIMO and OFDM systems, see e.g., \cite{R:WR1},\cite{R:MISO},\cite{R:OFDMA} and \cite{R:OFDM}. In \cite{R:WR1}, a three node MIMO broadcasting
 channel with separate energy harvesting and information decoding receiver was studied, where the rate-energy bound and region were derived for both TS and PS based schemes. In \cite{R:MISO}, the transmit beamforming at the multi-antenna base station and the PS strategy at the single-antenna users were jointly optimized for the MISO multi-user system. In \cite{R:OFDMA}, the authors optimized the PS for downlink OFDM system towards the objective of maximizing the bits/Joule energy efficiency, and in \cite{R:OFDM}, the authors investigated the optimal design of SWIET to maximize the weighted sum-rate for multi-user OFDM systems, where both TS and PS architectures were considered.
Nevertheless, these works just discussed the SWIET for MIMO systems and OFDM systems separately, which did not inherit
 the benefits of MIMO and OFDM for SWIET in the single system and the MIMO-OFDM channel was not considered. What's more, in these works, only point-to-point communication was considered and no relaying was involved.

\subsection{Motivation}
 So far, quite a few works have been done for MIMO-OFDM systems, especially for MIMO-OFDM relaying systems. For instance,
 \cite {R:RSB} studied the secure relay beamforming for SWIET in two-hop non-regenerative relay systems, where however, only the relay was deployed with multiple antennas. The source, the legitimate destination, the power receiver and the eavesdropper were all assumed with single antenna and non OFDM channel was considered.

To the best of the our knowledge, only two papers (see \cite{R:WP2} and \cite{R:WRb}) thus far have studied the SWIET for MIMO-OFDM relaying systems. In their work, a two-hop relaying MIMO-OFDM system was considered, where the source and relay were assumed to be two energy-supplied nodes and the destination was composed of one information receiver and one energy receiver. The energy receiver could harvest energy from the signals transmitted by both the source and the relay, and the information receiver can only collect the information from the signals forwarded by the relay. For such a two-hop MIMO-OFDM relay system, the authors studied its optimum performance boundaries and measured the rate-energy regions.

In this paper, we also focus on the SWIET for two-hop MIMO-OFDM relaying systems. We consider such a
scenario, where a source with fixed energy supply desires to transmit its information to a destination. Due to the
barrier between the source and the destination or the large distance over their direct link, the source cannot directly transmit its signal to the destination, so it asks a relay to assist its information transmission. However, because of the selfish nature (or lack of energy-supply), the relay is not willing to consume its own energy (or has no available energy) to help the source to forward the information. In this case, by employing the function of energy harvesting over RF signals, the relay is capable of harvesting energy from the wireless signals transmitted by the source and then uses its harvested energy to help the information delivering from the source to the destination. This scenario can be potentially applied in various energy-constrained networks. For example, in wireless sensor networks, where a sink node with energy supply wants to send its data (e.g., management data or signaling data) to the sensors which are too far away from it. So the sink node needs the help of some intermediary sensor node. But due to the  limited battery capacity, the intermediary sensor is not willing to help the sink node. For such an application, the SWIET can be employed to encourage the intermediary sensor to use the harvested energy to help the data transmission. For another example, one source station wants to transmit information to its destination station. There is a hill located between the two stations, so that they cannot communicate with each other. For such an application, a SWIET-aware relay station which has no fixed power supply due to the rugged environment can be deployed on the top of the hill or in the tunnel of the hill to help the information transmission from the source station to the destination station.

\subsection{Contributions}
Compared with previous works, some distinct features of our work are stressed here. In existing works (e.g., \cite{R:EP2,R:EP3,R:D2,R:HC,R:WR1,R:WP2,R:WRb}), both source and relay were assumed with fixed power supply and the energy harvesting function was employed at the destination node and their goal was to explore the performance region (rate-energy region) or trade-off for the SWIET, where however, how to efficiently use the harvested energy in the same single system was not considered. Thus, their considered systems just can be referred to as the ``\textbf{harvest-only}'' system. Whereas, in our work, we investigate the SWIET in a two-hop relaying system and the energy harvesting function is employed at the relay node, where all the harvested energy at the relay is used for helping the information transmission from the source to the destination. Our investigated system therefore can be referred to as the ``\textbf{harvest-and-use}'' EH system.

The contributions of this work are summarized as follows.
\textit{Firstly}, this is the first work on investigating the ``harvest-and-use'' SWIET two-hop MIMO-OFDM non-generating relay system, where both EH and the consumption of the harvested energy are jointly considered in a single system from a systematic perspective. For this, by adopting the two practical receiver architectures presented in \cite{R:EP1}, we design two protocols, TS-based MIMO-OFDM relaying protocol (TSR) and  PS-based MIMO-OFDM relaying protocol (PSR).
\textit{Secondly}, to explore the system performance limits of our proposed TSR and PSR, we mathematically formulate two optimization problems for them. The objective is to maximize the E2E achievable information rate via joint resource allocation. Specifically, for TSR, the source power allocation, the time switching factor, the subchannel paring between the two hops and the power assignment of the harvested energy at the relay node are jointly considered and optimized, and for PSR, the source power allocation, the power splitting factors, the subchannel paring over the two hops and the power assignment of the harvested energy at the relay node are jointly considered and optimized.
\textit{Thirdly}, since the two optimization problems are difficult to solve by directly using traditional methods. We adopt a decomposition process to each of them. By doing so, for TSR, we derive the closed-form result on the optimal energy transfer and some closed-form solutions on the conditional optimal power allocation at the source and the relay. Moreover, for high signal-to-noise ratio (SNR) case, with some approximating operations, we also present the closed-form solutions on the joint power allocation at the source and the relay. Based on these results, we design two optimization schemes for TSR, where one adopts an iterative manner in terms of the conditional optimal power allocation to find the final joint optimal power allocation at the source and the relay, and the other adopts the approximating joint power allocation. For PSR, we derive the closed-form result on the power splitting factors and design an efficient algorithm to find the joint optimal power allocation at the source and power splitting at the relay.
\textit{Finally}, we provide extensive numerical results to confirm our theoretical analysis of the proposed PSR and TSR. It is shown that both the optimized PSR and the optimized TSR can achieve performance gain compared with those with only  simple resource allocation and system configuration. It is also shown that the relay position greatly affects the performance of PSR and TSR protocols in terms of achievable information rate. Specifically, the proposed PSR and TSR show very different reactions to the variation of relay position. The achievable information rate of the optimized PSR monotonically decreases when the relay moves from the source towards the destination and that of the optimized TSR firstly decreases and then increases with the increment of source-relay distance and the relatively worse performance is obtained when the relay is placed in the middle of the source and the destination. This is the first time to observe such a phenomenon. The simulation results also show that the optimized PSR always outperforms the optimized TSR in the two-hop non-regenerative MIMO-OFDM system. Moreover, it is also indicated that increasing either the number of antennas or the number of subcarriers can bring system performance gain to both TSR and PSR.

The rest of the paper is organized as follows. Section II describes the system model, where the optimal structure for the two-hop non-regenerative MIMO-OFDM system is adopted. Section III presents the proposed TSR and PSR protocols on the basis of the  optimal structure described in Section II and then formulate an optimization problem for each of them. Section IV and V discuss how to optimize the proposed TSR and PSR, respectively. Section VI presents some numerical results to discuss the system performance of our optimized PSR and TSR. Finally, Section VII summarizes this work.

\emph{Notations}: The lower and upper case bold face letters, e.g., \textbf{x} and \textbf{X}, are used to represent
a column vector  \textbf{x} and a matrix \textbf{X}, respectively. $\textbf{X}^H$ denotes the complex conjugate transpose of \textbf{X}. \textbf{I} and \textbf{0} are used to denote an identity matrix and  all-zero
vector with appropriate dimensions, respectively.
$\|\textbf{x}\|$ denotes the Euclidean norm of a complex vector \textbf{x}. \textbf{X} $\sim$ $\mathcal{CN}(\bm{\mu},\bm{\Sigma})$ denotes
the elements of \textbf{X} following complex Gaussian distribution with
mean $\bm{\mu}$ and covariance $\bm{\Sigma}$.  $\mathbb{E}[\textbf{X}]$ denotes
the statistical expectation of matrix \textbf{X}. $\mathbb{C}^{x \times y}$ denotes the space of
$x \times y$ matrices with complex entries. For a square matrix \textbf{X}, tr(\textbf{X}), $\mid\textbf{X}\mid$, $\textbf{X}^{-1}$, and $\textbf{X}^{\frac{1}{2}}$ denote its trace, determinant, inverse, and square-root, respectively. $\textbf{X}\succeq 0$ means that S is positive definite. diag$\{x_1,..., x_M\}$ denotes an
$M\times M$ diagonal matrix with $x_1,..., x_M$ being its diagonal elements. Rank(\textbf{X}) denotes the rank of matrix \textbf{X} and $[x]^+$ means $\max\{0,x\}$. $x^*$ and $x^\sharp$ denote the optimized result and the conditionally optimized result of variable $x$, respectively. To simplify the expressions, we first summarize some commonly used symbols throughout the paper in Table \ref{TAB}.

\begin{table}[*htp!]
\caption{Symbol Notations}\label{TAB}
\centering
\begin{tabular}{c|l}
  \hline Notation & \quad\quad\quad\quad\quad Representation\\
  \hline
  ${\mathfrak{B}}$:&the total system bandwidth;\\
  $\mathcal{P}_{\textrm{S}}$:&the available transmit power at source;\\
  $\mathcal{P}_{\textrm{R}}$:&the available transmit power at relay;\\
  ${\textbf{H}}_{q}$:& the channel matrix of the channel with transmit node $q$;\\
  $\bm{s}_i$: &the source symbol vector at ${\rm S}$ over subcarrier $i$ with\\ &$\mathbb{E}[\bm{s}_i\bm{s}_i^H] = \bm{\textrm{I}}_{N_{\textrm{S}}}$;\\
  $\bm{F}_{q,i}$: &the processing matrix at node $q$ over subcarrier $i$;\\
  $\bm{B}_{i}$: &the processing gain matrix at ${\rm R}$ associated with subcarrier $i$;\\ $\bm{z}_{\textrm{q},i}$: &the received AWGN at node $q$ over subcarrier $i$;\\
  $\hat{\bm{x}}_i$: &the transmit information signal vector\\
  &over subcarrier $i$;\\
  $\bm{x}_i$: &the transmitted signal vector at ${\rm S}$ for\\
  &energy transfer over subcarrier $i$;\\
  $\bm{X}_{i}$: &the covariance matrix of $\bm{x}_i$, i.e., $\bm{X}_{i}=\mathbb{E}\{\bm{x}_i\bm{x}_i^H\}$;\\
  $\bm{\mathcal {X}}_{\textrm{S}}$: &the covariance  matrix $\bm{\mathcal {X}}_{\textrm{S}}=\{\bm{X}_{1},\bm{X}_{2},\cdot\cdot\cdot,\bm{X}_{K}\}$ at ${\rm S}$\\ &for the energy transfer in TSR;\\
  $p^{(i)}_{\textrm{\tiny S},n}$: & the transmit power at the source over the $i$-th subcarrier on\\ the &$n$-th spatial subchannel;\\
  $p^{(j)}_{\textrm{\tiny R},n'}$: & the transmit power at the relay over the $j$-th subcarrier on\\ the &$n'$-th spatial subchannel;\\
  $\omega_{\ell}$: &the power allocation factor at ${\rm S}$ over the $\ell$-th subchannel of\\ &the first hop for both TSR and PSR;\\
  $\varpi_{\ell'}$: &the power allocation factor at ${\rm R}$ over the $\ell'$-th subchannel of\\ &the second hop for TSR;\\
  $\theta_{\ell,\ell'}$: &the subchannel pairing indicator of the $\ell$-th subchannel of the\\
  &first hop and the $\ell'$-th subchannel of the first hop;\\
  $\alpha$: &the time switching factor for TSR;\\
  $\rho_{\ell}$: &the power splitting factor at ${\rm R}$ over the $\ell$-th E2E subchannel\\
  &for PSR;\\
  \hline
\end{tabular}
\end{table}

\section{System Model} \label{Sec:SectII}

\begin{figure}
\centering
\includegraphics[width=0.45\textwidth]{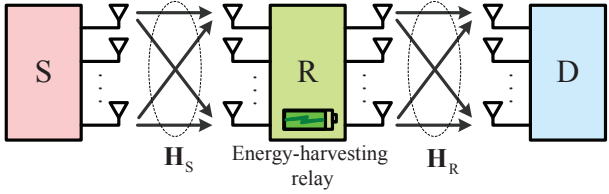}
\caption{System model of the two-hop MIMO-OFDM relay network with an energy-harvesting relay node, where $\textbf{H}_{q}=\{\bm{H}_{q,1},\bm{H}_{q,2},...,\bm{H}_{q,K}\}$ and $q\in\{\textrm{S},\textrm{R}\}$.}\label{Fig:model}
\end{figure}

Let us consider a two-hop relay network model, as shown in Figure  \ref{Fig:model},
where source ${\rm S}$ wants to transmit its information to destination ${\rm D}$.
Due to the long distance or the shielding effect caused by some barrier between ${\rm S}$ and ${\rm D}$, ${\rm D}$ is not within the communication range of ${\rm S}$, so that all signals received at ${\rm D}$ need to be forwarded by the assisting relay ${\rm R}$. Such a relaying model has been widely adopted to extend the communication coverage, which is well-known as the Type-II relaying in IEEE and 3GPP standards \cite{R:Relay}.

To achieve the benefits of MIMO technique, we assume that all nodes in the system are equipped with multiple antennas, where $N_{\rm S}$, $N_{\rm R}$ and $N_{\rm D}$ antennas are deployed at the source, the relay and the destination, respectively. Half-duplex mode and non-regenerative relaying are employed at ${\rm R}$ so that the signal transmission over the two hops are divided into two phases, i.e., a source phase and a relay phase, are involved in completing each round of information transmission from ${\rm S}$ to ${\rm D}$, where in the source phase, ${\rm S}$ transmits its signals  to ${\rm R}$, and in the relay phase,
${\rm R}$ amplifies the received signals and then forwards them  to ${\rm D}$.

With the deployment of OFDM, the total system bandwidth $\mathfrak{B}$ (i.e., frequency-selective channel) is divided into $K$ frequency-flat subcarriers (subchannels). Block fading channel model is assumed, so the channel gain over each subcarrier can maintain constant during each round of two-hop relaying transmission. To explore the potential capacity and performance limit of such a MIMO-OFDM non-regenerative relaying system, we assume that all nodes have full knowledge of the channel state information (CSI) and perfect synchronization over the two hops.

The source {\rm S} is with fixed energy supply while relay ${\rm R}$ is an energy-constrained/energy-selfish node with EH function deployed. That is, ${\rm R}$ has no energy (or is not willing to consume its own energy) to help ${\rm S}$ forward information but it is able to harvest energy from the transmitted signals of ${\rm S}$. Thus, ${\rm R}$ can use the harvested energy to help ${\rm S}$ transmit the information to ${\rm D}$. Denote $\mathcal{P}_{\textrm{S}}$ and $\mathcal{P}_{\textrm{R}}$ to be the average available power at ${\rm S}$ and average harvested power at ${\rm R}$, respectively. So, during each round of two-hop relaying, ${\rm S}$ first consumes $\mathcal{P}_{\textrm{S}}$ power to simultaneously transmit its energy and information to ${\rm R}$. Then, ${\rm R}$ uses up the harvested $\mathcal{P}_{\textrm{R}}$ power to help $\rm S$ transmit the information to ${\rm D}$ in the same round of time.

In the source phase, ${\rm S}$ transmits signals to ${\rm R}$. Let $\bm{s}_i\in \mathbb{C}^{N_{\textrm{S}}\times 1}$ be the source symbol vector to be transmitted at source ${\rm S}$ over subchannel $i$ from ${\rm S}$ to ${\rm R}$. $\mathbb{E}[\bm{s}_i\bm{s}_i^H] = \bm{\textrm{I}}_{N_{\textrm{S}}}$.
If we denote the channel matrix from ${\rm S}$ to ${\rm R}$ over subchannel $i$ as $\bm{H}_{\textrm{S},i}$,  the received signal vector at ${\rm R}$ over the $i$-th subcarrier in the source phase can be expressed as
\begin{equation}
\bm{y}_{\textrm{R},i}=\bm{H}_{\textrm{S},i}\bm{F}_{\textrm{S},i}\bm{s}_i+\bm{z}_{\textrm{R},i},\,\,i\in\{1,2,...,K\}
\end{equation}
where $\bm{z}_{\textrm{R},i}\sim \mathcal {CN}(0,\sigma_{\textrm{R}}^2\bm{\textrm{I}}_{N_{\textrm{R}}})$ represents the received noise at ${\rm R}$ over subcarrier $i$. $\bm{F}_{\textrm{S},i}\in \mathbb{C}^{N_{\textrm{S}}\times N_{\textrm{S}}}$ is the precoding matrix at ${\rm S}$.

In the relay phase, ${\rm R}$ amplifies the received signals over subcarrier $i$ of the first hop and then forwards them to ${\rm D}$ over subcarrier $j$ ($j\in\{1,2,...,K\}$) of the second hop. Note that, without using subcarrier-pairing, the received signal at ${\rm R}$ over subcarrier $i$ is also forwarded to ${\rm D}$ over subcarrier $i$, i.e., $i=j$, while if subcarrier-paring is employed, $j$ may be different from $i$. For clarity, we use $(i,j)$ to represent a subcarrier pair composed of the $i-$th subcarrier over the $\rm{S}-\rm{D}$ link and the $j-$th subcarrier over the $\rm{R}-\rm{D}$ link. Let $\bm{H}_{\textrm{R},j}$ and $\bm{z}_{\textrm{D},j}\sim \mathcal {CN}(0,\sigma_{\textrm{D}}^2\bm{\textrm{I}}_{N_{\textrm{D}}})$ be the channel matrix from ${\rm R}$ to ${\rm D}$ and the additive noise received at ${\rm D}$ over subcarrier $j$, respectively. Then, the received signal at ${\rm D}$ over the $(i,j)$ subcarrier pair can be given by
\begin{flalign}\label{Eq:yd}
\bm{y}_{\textrm{D},j}&=\bm{F}_{\textrm{D},j}(\bm{H}_{\textrm{R},j}\bm{F}_{\textrm{R},i}\bm{y}_{\textrm{R},i}+\bm{z}_{\textrm{D},j})\\
=&\bm{F}_{\textrm{D},j}(\bm{H}_{\textrm{R},j}\bm{F}_{\textrm{R},i}(\bm{H}_{\textrm{S},i}\bm{F}_{\textrm{S},i}\bm{s}_i+\bm{z}_{\textrm{R},i}))+\bm{F}_{\textrm{D},j}\bm{z}_{\textrm{D},j}\nonumber\\
=&\bm{F}_{\textrm{D},j}(\bm{H}_{\textrm{R},j}\bm{F}_{\textrm{R},i}\bm{H}_{\textrm{S},i}\bm{F}_{\textrm{S},i}\bm{s}_i+\bm{H}_{\textrm{R},j}\bm{F}_{\textrm{R},i}\bm{z}_{\textrm{R},i})+\bm{F}_{\textrm{D},j}\bm{z}_{\textrm{D},j},\nonumber
\end{flalign}
where $\bm{F}_{\textrm{R},i}\in \mathbb{C}^{N_{\textrm{R}}\times N_{\textrm{R}}}$ and $\bm{F}_{\textrm{D},j}\in \mathbb{C}^{N_{\textrm{D}}\times N_{\textrm{D}}}$ denote the forwarding matrix at ${\rm R}$ and the processing matrix at ${\rm D}$, respectively.

Since each node knows the full CSI, the singular
value decomposition (SVD) of all channel matrices is available for the system to determine
the transmit-and-receive processing matrices at all nodes. The SVD of the channel matrices over the two hops can be expressed by
\begin{flalign}\label{Eq:SVD}
\left\{ \begin{aligned}
 &\bm{H}_{\textrm{S},i}=\bm{U}_{\textrm{S},i}\bm{\Lambda}_{\textrm{S},i}\bm{V}_{\textrm{S},i}^H,\\
 &\bm{H}_{\textrm{R},j}=\bm{U}_{\textrm{R},j}\bm{\Lambda}_{\textrm{R},j}\bm{V}_{\textrm{R},j}^H,\\
 \end{aligned}\right.
\end{flalign}
where  $\bm{\Lambda}_{\textrm{S},i} \in \mathbb{C}^{N_{\textrm{R}}\times N_{\textrm{S}}}$ and $\bm{\Lambda}_{\textrm{R},j}\in \mathbb{C}^{N_{\textrm{D}}\times N_{\textrm{R}}}$ are two diagonal
matrices with non-negative real numbers on the diagonal. $\bm{\Lambda}_{\textrm{S},i}=\textrm{diag}\Big\{\sqrt{\lambda^{(i)}_{\textrm{\tiny S},1}},\sqrt{\lambda^{(i)}_{\textrm{\tiny S},2}},\cdot\cdot\cdot,\sqrt{\lambda^{(i)}_{\textrm{\tiny S},\textrm{Rank}(\bm{H}_{\textrm{S},i})}}\,\Big\}$ with $\sqrt{\lambda^{(i)}_{\textrm{\tiny S},1}}\geq\sqrt{\lambda^{(i)}_{\textrm{\tiny S},2}}\geq\cdot\cdot\cdot\geq\sqrt{\lambda^{(i)}_{\textrm{\tiny S},\textrm{Rank}(\bm{H}_{\textrm{S},i})}}$ and $\bm{\Lambda}_{\textrm{R},j}=\textrm{diag}\Big\{\sqrt{\lambda^{(j)}_{\textrm{\tiny R},1}},\sqrt{\lambda^{(j)}_{\textrm{\tiny R},2}},\cdot\cdot\cdot,\sqrt{\lambda^{(j)}_{\textrm{\tiny R},\textrm{Rank}(\bm{H}_{\textrm{R},j})}}\,\Big\}$ with $\sqrt{\lambda^{(j)}_{\textrm{\tiny R},1}} \geq \sqrt{\lambda^{(j)}_{\textrm{\tiny R},2}}\geq \cdot\cdot\cdot \geq \sqrt{\lambda^{(j)}_{\textrm{\tiny R},\textrm{Rank}(\bm{H}_{\textrm{R},j})}}$. $\bm{U}_{\textrm{S},i} \in \mathbb{C}^{N_{\textrm{R}}\times N_{\textrm{R}}}$, $\bm{V}_{\textrm{S},i} \in \mathbb{C}^{N_{\textrm{S}}\times N_{\textrm{S}}}$, $\bm{U}_{\textrm{R},j} \in \mathbb{C}^{N_{\textrm{D}}\times N_{\textrm{D}}}$ and $\bm{V}_{\textrm{R},j}\in \mathbb{C}^{N_{\textrm{R}}\times N_{\textrm{R}}}$ are four  complex unitary
matrices, so they do not change the
statistics of the channel. This implies that by using SVD, the mutual
information of the corresponding channels can be preserved \cite{R:MIMOOFDM}.

Moreover, it was proved that by performing the SVD on $\bm{H}_{\textrm{S},i}$ and $\bm{H}_{\textrm{R},i}$, the achievable rate of the two-hop non-regenerative MIMO channel can be maximized when the overall two-hop channel is decomposed into a number of parallel uncorrelated paths \cite{R:MIMOAF}. We therefore adopt such a formulation to design the SWIET for the two-hop non-regenerative MIMO-OFDM relay channel as follows.

Substituting (\ref{Eq:SVD}) into (\ref{Eq:yd}), it can be obtained that
\begin{flalign}\label{Eq:ydn}
\bm{y}_{\textrm{D},j}
\\
=&\bm{F}_{\textrm{D},j}\bm{H}_{\textrm{R},j}\bm{F}_{\textrm{R},i}\bm{H}_{\textrm{S},i}\bm{F}_{\textrm{S},i}\bm{s}_i+\bm{F}_{\textrm{D},j}\bm{H}_{\textrm{R},j}\bm{F}_{\textrm{R},i}\bm{z}_{\textrm{R},i}+\bm{F}_{\textrm{D},j}\bm{z}_{\textrm{D},j}
\nonumber\\
=&\bm{F}_{\textrm{D},j}\bm{U}_{\textrm{R},j}\bm{\Lambda}_{\textrm{R},j}\bm{V}_{\textrm{R},j}^H\bm{F}_{\textrm{R},i}\bm{U}_{\textrm{S},i}\bm{\Lambda}_{\textrm{S},i}\bm{V}_{\textrm{S},i}^H\bm{F}_{\textrm{S},i}\bm{s}_i
\nonumber\\
&+\bm{F}_{\textrm{D},j}\bm{U}_{\textrm{R},j}\bm{\Lambda}_{\textrm{R},j}\bm{V}_{\textrm{R},j}^H\bm{F}_{\textrm{R},i}\bm{z}_{\textrm{R},i}
+\bm{F}_{\textrm{D},j}\bm{z}_{\textrm{D},j}.
\nonumber
\end{flalign}
Assume that the allocated power at ${\rm S}$ over subcarrier $i$ is ${P}_{\textrm{S},i}$ satisfying that $\sum\nolimits_{i=1}^{K} {P}_{\textrm{S},i} \leq \mathcal{P}_{\textrm{S}}$. Then, we have that
\begin{flalign}\label{Eq:ydFS}
\bm{F}_{\textrm{S},i}=\sqrt{{P}_{\textrm{S},i}}\bm{V}_{\textrm{S},i}\textbf{w}_{\textrm{s}}^{(i)},
\end{flalign}
where $\textbf{w}_{\textrm{s}}^{(i)}\triangleq \textrm{diag}\Big\{\sqrt{\textit{w}_{s,1}^{(i)}},\sqrt{\textit{w}_{s,2}^{(i)}},...,\sqrt{\textit{w}_{s,N_{\textrm{S}}}^{(i)}}\Big\}$ is a diagonal matrix composed of weighting coefficients of the antennas at ${\rm S}$ over the $i-$th subcarrier.
Then the transmit information signal vector from the source over subcarrier $i$ thus can be given by $\hat{\bm{x}}_i=\sqrt{{P}_{\textrm{S},i}}\bm{V}_{\textrm{S},i}\textbf{w}_{\textrm{s}}^{(i)}\bm{s}_i$.
$\bm{F}_{\textrm{D},j}$ and $\bm{F}_{\textrm{R},i}$ are chosen as $\bm{F}_{\textrm{D},j}=\bm{U}_{\textrm{R},j}^H$ and $\bm{F}_{\textrm{R},i}=\bm{V}_{\textrm{R},i}\bm{B}_{i}\bm{U}_{\textrm{S},i}^H$, in order to obtain the parallel single-input single-output (SISO) paths.
Such a design of $\bm{F}_{\textrm{R},i}$ implies a linear processing, which was also adopted in some existing works for two-hop amplified-and-forward MIMO systems (e.g.,\cite{R:MIMOAF}).
For simplicity, the parallel SISO paths are referred to as end-to-end (E2E) subchannels in the sequel. Note that, for a given subcarrier pair ($i,j$), $\bm{F}_{\textrm{R},i}$ and $\bm{F}_{\textrm{R},j}$ actually represent the same processing matrix at $\texttt{R}$, i.e., $\bm{F}_{\textrm{R},i}=\bm{F}_{\textrm{R},j}$, resulting in  $\bm{V}_{\textrm{R},i}=\bm{V}_{\textrm{R},j}$. Therefore, substituting (\ref{Eq:ydFS}) and $\bm{F}_{\textrm{R},i}=\bm{V}_{\textrm{R},j}\bm{B}_{i}\bm{U}_{\textrm{S},i}^H$ into (\ref{Eq:ydn}) yields
\begin{flalign}\label{Eq:ydnni}
\bm{y}_{\textrm{D},j}=\bm{\Lambda}_{\textrm{R},j}\bm{B}_{i}\bm{\Lambda}_{\textrm{S},i}\hat{\bm{x}}_i
+\bm{\Lambda}_{\textrm{R},j}\bm{B}_{i}\bm{U}_{\textrm{S},i}^H\bm{z}_{\textrm{R},i}+\bm{U}_{\textrm{R},j}^H\bm{z}_{\textrm{D},j}.
\end{flalign}
where $\bm{B}_{i}$ can be regarded as a processing gain matrix at ${\rm R}$.

With above-mentioned operations, one can see that both the $i$-th subcarrier of the first hop and the $j$-th subcarrier of the second hop are divided into some orthogonal spatial subchannels, and with the one-to-one concatenation between the spatial subchannels over the two hops, a set of orthogonal E2E subchannels are obtained.
As a result, it can be deduced that the number of the E2E subchannels must be bounded to the minimum number of spatial subchannels of each single hop. This means that in a two-hop non-regenerative MIMO-OFDM relaying, for each subcarrier, only a subset of spatial subchannels either in the first or the second hop is used. Let $N$ be the dimension of the subset. It can be inferred that
\begin{flalign}
N=\min\{\textrm{Rank}(\bm{\bm{H}_{\textrm{S},i}), \textrm{Rank}(H}_{\textrm{R},j})\}=\min\{N_{\textrm{S}}, N_{\textrm{R}}, N_{\textrm{D}}\}.
\end{flalign}
Let $\hat{\bm{U}}_{q,i}$, $\hat{\bm{\Lambda}}_{q,i}$ and $\hat{\bm{V}}_{q,i}$ be $N\times N$ matrices, which are composed of the first $N$ columns and $N$ rows of $\bm{U}_{q,i}$, $\bm{\Lambda}_{q,i}$ and $\bm{U}_{q,i}$, respectively, where $q\in\{\textrm{S},\textrm{R}\}$.
Then,
(\ref{Eq:ydnni}) is equivalently transformed to be
\begin{flalign}\label{Eq:ydnn}
\bm{y}_{\textrm{D},j}=\hat{\bm{\Lambda}}_{\textrm{R},j}\bm{B}_{i}\hat{\bm{\Lambda}}_{\textrm{S},i}\hat{\bm{x}}_i
+\hat{\bm{\Lambda}}_{\textrm{R},j}\bm{B}_{i}\hat{\bm{U}}_{\textrm{S},i}^H\bm{z}_{\textrm{R},i}+\hat{\bm{U}}_{\textrm{R},j}^H\bm{z}_{\textrm{D},j}.
\end{flalign}
Thus, the mutual information over the $(i,j)$ subcarrier pair for the non-regenerative MIMO-OFDM system can be given by
\begin{flalign}\label{Eq:MI}
I(\hat{\bm{x}}_i,\hat{\bm{y}}_j)=\frac{\mathfrak{B}}{2K}\log \Bigg|\textbf{I}_N+ \frac{\hat{\bm{\Lambda}}_{\textrm{R},j}\bm{B}_{i}\hat{\bm{\Lambda}}_{\textrm{S},i}\bm{\Delta}_i\hat{\bm{\Lambda}}_{\textrm{S},i}^H\bm{B}_{i}^H\hat{\bm{\Lambda}}_{\textrm{R},j}^H}
{\sigma_{\textrm{R}}^2\hat{\bm{\Lambda}}_{\textrm{R},j}\bm{B}_{i}\bm{B}_{i}^H\hat{\bm{\Lambda}}_{\textrm{R},j}^H+\sigma_{\textrm{D}}^2\textbf{I}}\Bigg|,
\end{flalign}
where $\mathfrak{B}$ is the total bandwidth of the OFDM system and $\frac{1}{2}$ is used to describe the time division feature of the two-hop retransmission. $\bm{\Delta}_i=\mathbb{E}\{\hat{\bm{x}}_i\hat{\bm{x}}_i^H\}$.

From (\ref{Eq:MI}), it can be seen that each subcarrier pair $(i,j)$ is divided into $N$ effective E2E subchannels. As there are $K$ subcarriers in the MIMO-OFDM system, the total $KN$ E2E subchannels can be obtained by using the SVD decomposition over the two hops. In order to maximize $I(\hat{\bm{x}}_i,\hat{\bm{y}}_j)$, the signal transmitted over subcarrier $i$ on the $n$-th spatial subchannel of the first hop is allowed to be forwarded by ${\rm R}$ over the $j$-th subcarrier on the spatial subchannel $n'$ of the second hop, this is referred to as subchannel pairing in the sequel. According to \cite{R:MIMOAF} and \cite{R:MIMOOFDM},  $\bm{B}_{i}$ satisfies that
\begin{flalign}\label{Eq:Gr}
\bm{B}_{i}=\textrm{diag}\big[\beta_{i,1},\beta_{i,2},\cdot\cdot\cdot,\beta_{i,N}\big],
\end{flalign}
with
\begin{flalign}\label{Eq:beta}
\beta_{i,n}=\sqrt{\tfrac{p^{(j)}_{\textrm{\tiny R},n'}}{p^{(i)}_{\textrm{\tiny S},n}\lambda^{(i)}_{\textrm{\tiny S},n}+\sigma_{\textrm{\tiny R}}^2}},
\end{flalign}
where $p^{(i)}_{\textrm{\tiny S},n}$, $p^{(j)}_{\textrm{\tiny R},n'}$ denote the transmit power of the source over the $i$-th subcarrier on the $n$-th spatial subchannel, the transmit power of the relay over the $j$-th subcarrier on the $n'$-th  spatial subchannel, respectively.

For notation simplification, we define $\ell \triangleq (i-1)K+n$ and $\ell' \triangleq (j-1)K+n'$, where $1\leq n, n'\leq N$ and $1\leq i,j\leq K$. Thus, it can be seen that that $1 \leq\ell\leq KN$ and $1\leq\ell'\leq KN$. Consequently, the subchannel pairing mentioned previously can be redescribed as the $\ell$-th subchannel of the first hop is paired with the $\ell'$-th subchannel of the second hop. Therefore,
(\ref{Eq:beta}) is rewritten as
\begin{flalign}\label{Eq:beta2}
\beta_{i,n}=\beta_{\ell}=\sqrt{\frac{p_{\textrm{\tiny R},\ell'}}{p_{\textrm{\tiny S},\ell}\lambda_{\textrm{\tiny S},\ell}+\sigma_{\textrm{\tiny R}}^2}},
\end{flalign}
where $p_{\textrm{\tiny S},\ell}\triangleq p^{(i)}_{\textrm{\tiny S},n}$, $p_{\textrm{\tiny R},\ell}\triangleq p^{(j)}_{\textrm{\tiny R},n'}$ and $\lambda_{\textrm{\tiny S},\ell}\triangleq \lambda^{(i)}_{\textrm{\tiny S},n}$.  Clearly, $\sum\nolimits_{\ell=1}^{KN} p_{\textrm{\tiny S},\ell} \leq \mathcal{P}_{\textrm{S}}$ and $\sum\nolimits_{\ell'=1}^{KN} p_{\textrm{\tiny R},\ell'} \leq \mathcal{P}_{\textrm{R}}$. By adopting the structure described above, which was presented to maximize the instantaneous E2E capacity in \cite{R:MIMOAF}, the achievable instantaneous information rate over the subchannel pair $(\ell,\ell')$ can be given by
\begin{flalign}\label{Eq:R}
R_{\ell,\ell'}=\frac{\mathfrak{B}}{2K}\log\bigg(1+\frac{p_{\textrm{\tiny S},\ell}p_{\textrm{\tiny R},\ell'}\frac{\lambda_{\textrm{\tiny S},\ell}}{\sigma_{\textrm{\tiny R}}^2}\frac{\lambda_{\textrm{\tiny R},\ell'}}{\sigma_{\textrm{\tiny D}}^2}}
{1+p_{\textrm{\tiny S},\ell}\frac{\lambda_{\textrm{\tiny S},\ell}}{\sigma_{\textrm{\tiny R}}^2}+p_{\textrm{\tiny R},\ell'}\frac{\lambda_{\textrm{\tiny R},\ell'}}{\sigma_{\textrm{\tiny D}}^2}}\bigg).
\end{flalign}

\section{Protocols and Optimization Problem Formulation}

Based on the structure described in  Section \ref{Sec:SectII}, in this section we shall present two protocols for the two-hop non-regenerative MIMO-OFDM  system by considering the two practical receiver architectures proposed in \cite{R:EP1}. Moreover, to explore the system performance limit, we also formulate two optimization problems for the two protocols in this Section.

\subsection{Protocol Description and Optimization Problem Formulation for TSR}\label{Sec:TSRI}
\subsubsection{TSR Protocol}\label{Sec:TSR}
Firstly, we consider the TS receiver architecture at ${\rm R}$ and then propose a time switching-based EH non-regenerative MIMO-OFDM Relaying (TSR) protocol as follows.

\begin{figure}
\centering
\includegraphics[width=0.49\textwidth]{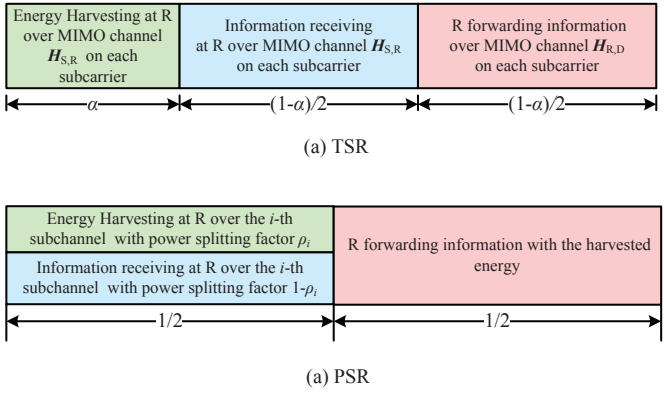}
\caption{The framework of our proposed two protocols: (a) TSR and (b) PSR.}\label{Fig:Framework}
\end{figure}

The framework of TSR is illustrated in Figure \ref{Fig:Framework}(a), in which each time period $T$ is divided into three phases. The first phase is assigned with a duration of $\alpha T$, which is used for energy transfer from ${\rm S}$ to ${\rm R}$. $\alpha \in [0,1]$ denotes the \textit{time switching factor}. The second and the third phases are assigned with equal time duration of $\frac{1-\alpha}{2}T$, where the second phase is used for the information transmission from ${\rm S}$ to ${\rm R}$ and the third one is used for ${\rm R}$ to forward the received information to ${\rm D}$.

In the first phase of TSR, only energy is transferred. As the EH receiver does not need to convert the received signal from the RF band to harvest the carried energy,  in order to seek a better system performance, energy transfer can be different from information transmission.  Denote the transmitted signal vector at ${\rm S}$ for energy transfer over subcarrier $i$ to be $\bm{x}_i$. $\bm{x}_i$ may be different from the information symbol vector $\bm{\hat{x}}_i$ defined in Section \ref{Sec:SectII}. Thus, the received signal at ${\rm R}$ for energy harvesting can be given by
\begin{equation}
\bm{y}_{{\textrm{R},i}}^{\textrm{(EH)}}=\bm{H}_{\textrm{S},i}\bm{x}_i+\bm{z}_{\textrm{R},i},
\end{equation}
Similar to some existing works (see e.g., \cite{R:OFDM}), we also assumed that the total harvested RF-band energy at ${\rm R}$ over the subcarrier $i$ is proportional to that of received baseband signal. Without loss of generality, we normalize the time period $T$ to 1 hereafter in this paper. Then the total harvested RF-band energy at ${\rm R}$ over the subcarrier $i$ can be expressed by
\begin{flalign}
E_{\textrm{R},i}&=\alpha \eta\parallel\bm{H}_{\textrm{S},i}\bm{x}_i\parallel^2,
\end{flalign}
where $\eta$ is a constant, which is used to describe the converting efficiency of the energy transducer in converting the harvested energy to electrical energy to be stored. In this paper, we assume $\eta=1$. Note that such an assumption is also widely adopted in the exploration of system performance limit for the convenience of analysis, see e.g., \cite{R:WP1}.

Define $\bm{X}_{i}=\mathbb{E}\{\bm{x}_i\bm{x}_i^H\}$ as the covariance matrix of $\bm{x}_i$. The total transmit power over all $K$ subcarriers can be given by $\sum\nolimits_{i=1}^{K}\mathbb{E}\{\parallel\bm{x}_i\parallel^2\}=\sum\nolimits_{i=1}^{K}\textrm{tr}(\bm{X}_{i})$, which is constrained by the available power at ${\rm S}$,  i.e.,
\begin{flalign}
\sum\nolimits_{i=1}^{K}\textrm{tr}(\bm{X}_{i})\leq \mathcal{P}_{\textrm{S}},
\end{flalign}
where $\textrm{tr}(\bm{X}_{i})$ actually can be regarded the transmit power allocated over subcarrier $i$ for energy transfer.

In the second phase of TSR, ${\rm S}$ transmit the signals over all $\ell\in\{1,2,...,KN\}$ subchannels under the average power constraint $\mathcal{P}_{\textrm{S}}$ and in the third phase of TSR, ${\rm R}$ forward the received signals over all $\ell'\in\{1,2,...,KN\}$ subchannels  under the available power constraint $\mathcal{P}_{\textrm{R}}$ with the optimal structure described in Section \ref{Sec:SectII}. Since all energy harvested in the first phase is used for the information relaying, it can be deduced that the available transmit power at ${\rm R}$ for the information forwarding is
\begin{equation}\label{Eq:PR}
\mathcal{P}_{\textrm{R}}=\frac{\sum\nolimits_{i=1}^{K}E_{\textrm{R},i}}{\frac{(1-\alpha)}{2}}=\frac{2\alpha }{1-\alpha}\sum\nolimits_{i=1}^{K}{\parallel\bm{H}_{\textrm{S},i}\bm{x}_i\parallel^2}.
\end{equation}

Define $p_{\textrm{\tiny S},\ell}\triangleq \mathcal{P}_{\textrm{S}}\omega_{\ell}$
and $p_{\textrm{\tiny R},\ell'}\triangleq \mathcal{P}_{\textrm{R}}\varpi_{\ell'}$, where $0 \leq \omega_{\ell}\leq 1$ is used to represent the power allocation factor at ${\rm S}$ for subcarrier $i$ on the spatial subchannel $n$ (i.e., subchannel $\ell$ of the first hop) satisfying that $\sum\nolimits_{\ell=1}^{KN}{\omega_{\ell}}\leq 1$,  and $\varpi_{\ell'}\in[0,1]$ is used to represent the power allocating factor at ${\rm R}$ over subcarrier $j$ on spatial subchannel $n'$ (i.e., subchannel $\ell'$ of the second hop) satisfying that $\sum\nolimits_{\ell'=1}^{KN}{\varpi_{\ell'}}\leq 1$. Note that, since $\mathcal{P}_{\textrm{S}}\omega_{\ell}={P}_{\textrm{S},i}\textit{w}_{\textrm{s},n}^{(i)}$ and  ${P}_{\textrm{S},i}=\mathcal{P}_{\textrm{S}}\sum\nolimits_{\ell=(i-1)K+1}^{(i-1)K+N}{\omega_{\ell}}$,
it can be inferred that $\textit{w}_{\textrm{s},n}^{(i)}=\omega_{\ell}/\sum\nolimits_{b=(i-1)K+1}^{(i-1)K+N}{\omega_{b}}$, which means that if we determine $\omega_{\ell}$ for $\ell\in\{1,2,...,KN\}$, $\textit{w}_{\textrm{s},n}^{(i)}$ can also be determined for all $n\in\{1,2,...,N\}$ and $i\in\{1,2,...,K\}$. Therefore, we shall discuss how to optimize $\omega_{\ell}$ instead of $\textit{w}_{\textrm{s},n}^{(i)}$ in the sequel. Therefore, $R_{\ell,\ell'}$ in (\ref{Eq:R}) can be rewritten to be
\begin{flalign}\label{R:vvv}
R^{(\textrm{TSR})}_{\ell,\ell'}=\frac{\mathfrak{B}}{2K}\log\bigg(1+\frac{\mathcal{P}_{\textrm{S}}\omega_{\ell}\mathcal{P}_{\textrm{R}}\varpi_{\ell'}\frac{\lambda_{\textrm{\tiny S},\ell}}{\sigma_{\textrm{\tiny R}}^2}\frac{\lambda_{\textrm{\tiny R},\ell'}}{\sigma_{\textrm{\tiny D}}^2}}
{1+\mathcal{P}_{\textrm{S}}\omega_{\ell}\frac{\lambda_{\textrm{\tiny S},\ell}}{\sigma_{\textrm{\tiny R}}^2}+\mathcal{P}_{\textrm{R}}\varpi_{\ell'}\frac{\lambda_{\textrm{\tiny R},\ell'}}{\sigma_{\textrm{\tiny D}}^2}}\bigg)
\end{flalign}
for TSR.

Based on (\ref{R:vvv}), the total E2E instantaneous achievable information rate of the TSR is given by
\begin{flalign}\label{CTSR}
C_{\textrm{TSR}}=\frac{1-\alpha}{2}\sum\nolimits_{\ell=1}^{KN}\sum\nolimits_{\ell'=1}^{KN} {\theta_{\ell,\ell'}{R^{(\textrm{TSR})}_{\ell,\ell'}}},
\end{flalign}
where $\theta_{\ell,\ell'}\in\{0,1\}$ is the indicator of the subchannel-pairing. Specifically, $\theta_{\ell,\ell'}=1$ means that the $\ell$-th subchannel of the first hop is paired with the $\ell'$-th subchannel of the second hop. Otherwise, $\theta_{\ell,\ell'}=0$.

\subsubsection{Optimization Problem Formulation for TSR}\label{Sec:Problem}
From the description in Section \ref{Sec:TSR}, one can see that for the TSR, only the available transmit power at the source $\mathcal{P}_{\textrm{S}}$ and all channel matrices are known, while the rest parameters such as the covariance  matrix $\bm{\mathcal {X}}_{\textrm{S}}=\{\bm{X}_{1},\bm{X}_{2},\cdot\cdot\cdot,\bm{X}_{K}\}$ at ${\rm S}$ for the energy transfer, the time switching factor $\alpha$, the power allocating matrix $\bm{\omega}=\{\omega_{\ell}\}_{KN\times 1}$ at ${\rm S}$, the power allocating vector $\bm{\varpi}=\{\varpi_{\ell'}\}_{KN\times 1}$ at ${\rm R}$ and the subchannel pairing matrix $\bm{\theta}=\{\theta_{\ell,\ell'}\}_{KN \times KN}$ are all required to be determined and configured. Since all these parameters may affect the system performance, it is necessary to jointly design them for achieving the optimal performance. In this subsection, we formulate an optimization problem for it and we shall investigate how to solve the optimization problem in Section \ref{Sec:OPTDTSR}. Our goal is to find the optimal $\bm{\omega}^*$, $\bm{\varpi}^*$, $\bm{\theta}^*$ and $\alpha^*$ to maximize the end-to-end achievable information rate of TSR. The corresponding optimization problem can be mathematically expressed as
\begin{flalign}\label{Opt:TSR}
\mathop {\max }\limits_{\bm{\mathcal {X}}_{\textrm{S}},\bm{\omega},\bm{\varpi},\bm{\theta},\alpha} &C_{\textrm{TSR}}\\
\textrm{s.t.}\,\,&\sum\nolimits_{\ell=1}^{KN}{\omega_{\ell}}\leq 1,\,\,\sum\nolimits_{\ell'=1}^{KN}{\varpi_{\ell'}}\leq 1\nonumber\\
&\sum\nolimits_{\ell=1}^{KN}{\theta_{\ell,\ell'}}=1, \forall \ell';
\sum\nolimits_{\ell'=1}^{KN}{\theta_{\ell,\ell'}}=1, \forall \ell\nonumber\\
&\sum\nolimits_{i=1}^{K}\textrm{tr}(\bm{X}_{i})\leq \mathcal{P}_{\textrm{S}},\,\,\bm{X}_{i}\succeq 0,\nonumber\\
&0\leq \alpha \leq 1,\,\,\theta_{\ell,\ell'}\in\{0,1\}\nonumber
\end{flalign}
where the constraints $\sum\nolimits_{\ell=1}^{KN}{\omega_{\ell}}\leq 1$ and $\sum\nolimits_{\ell'=1}^{KN}{\varpi_{\ell}}\leq 1$ imply that the available transmit power at the source and the relay are limited by $\mathcal{P}_{\textrm{S}}$ and $\mathcal{P}_{\textrm{R}}$ in (\ref{Eq:PR}), respectively. The constraints $\sum\nolimits_{\ell=1}^{KN}{\theta_{\ell,\ell'}}=1, \forall \ell'$ and $\sum\nolimits_{\ell'=1}^{KN}{\theta_{\ell,\ell'}}=1, \forall \ell$ mean that each subchannel of the first hop is only allowed to be paired one subchannel of the second hop, vise verse. The constraint $\sum\nolimits_{i=1}^{K}\textrm{tr}(\bm{X}_{i})\leq \mathcal{P}_{\textrm{S}}$ indicates that the energy transfer in the first phase of TSR is also constrained by the available power at the source.

\subsection{Protocol Description and Optimization Problem Formulation for PSR}
\subsubsection{PSR Protocol}
In this subsection, we shall present another relaying protocol, power splitting-based EH non-regenerative MIMO-OFDM Relaying (PSR) by considering the PS receiver architecture presented in \cite{R:EP1}.

The framework of PSR is illustrated in Figure \ref{Fig:Framework}(b), in which the total time period $T$ is equally divided into two parts, where in the first $\frac{T}{2}$, energy and information are simultaneously transferred from ${\rm S}$ to ${\rm R}$ and in the rest $\frac{T}{2}$, ${\rm R}$ uses the harvested energy and forwards the received information to ${\rm D}$. Specifically, the receiver at ${\rm R}$ can be explained as follows. The received signal over the $\ell$-th subchannel at ${\rm R}$ is firstly corrupted by a
Gaussian noise at the RF-band, which is
assumed to have zero mean and equivalent baseband power. The RF-band signal is then fed into a power splitter,
which is assumed to be perfect without any noise induced.
After the power splitter, a portion of signal power is allocated to
the EH receiver. Suppose $\rho_{\ell}$ be the power splitting factor for the EH receiver. Then the rest $1-\rho_{\ell}$ part is input into the information receiver. The signal split to the information receiver then goes through
a sequence of non-regenerative relaying with the system structure described in Section \ref{Sec:SectII}.

\begin{figure}
\centering
\includegraphics[width=0.45\textwidth]{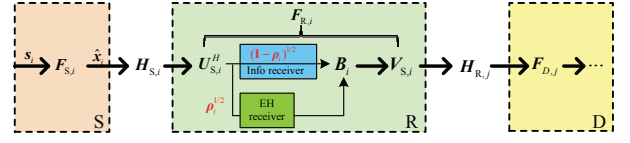}
\caption{The structure  of our proposed PSR.}\label{Fig:PSR}
\end{figure}

For clarity, the structure of our proposed PSR is illustrated in Figure \ref{Fig:PSR}, where at ${\rm R}$ the received signals are firstly processed by the matrix $\bm{U}_{\textrm{S},i}^H$ and then split into two flows with a power splitting factor matrix $\bm{\rho}$. After this, the $(\textbf{I}-\bm{\rho})$ part is input into the ``info receiver'' and the rest $\bm{\rho}$ part is input into the ``EH receiver''. The harvested energy by the EH receiver is then allocated to the information flow via the amplifying gain matrix $\bm{B}_i$. The detailed process is described as follows.

Let $\rho_{i,n}$ be the portion of the power split to the EH receiver over the $n$-th spatial subchannel of the $i$-th subcarrier at ${\rm R}$, $n \in \{1,2,..., N\}$. Thus, the total harvested RF-band energy at ${\rm R}$  over the subcarrier $i$ is proportional to that of received baseband signal, which can be expressed by
\begin{flalign}\label{Eq:psEr}
\mathcal {E}_{\textrm{R},i}&=\eta\parallel\bm{\rho}_i^{\frac{1}{2}}\bm{U}_{\textrm{S},i}^H\bm{H}_{\textrm{S},i}\bm{F}_{\textrm{S},i}\bm{s}_i\parallel^2,
\end{flalign}
where $\bm{\rho}_i=\textrm{diag}\{\rho_{i,1},\rho_{i,2},...,\rho_{i,N}\}$.
With the assumption of $\eta=1$, we have that
\begin{flalign}\label{Eq:psEr}
\mathcal {E}_{\textrm{R},i}=&\parallel\sqrt{{P}_{\textrm{S},i}}\bm{\rho}_i^{\frac{1}{2}}\bm{U}_{\textrm{S},i}^H\bm{H}_{\textrm{S},i}\bm{V}_{\textrm{S},i}\textbf{w}_{\textrm{s}}^{(i)}\bm{s}_i\parallel^2,\nonumber\\
=&{P}_{\textrm{S},i}\textrm{tr}(\bm{\rho}_i^{\frac{1}{2}}\bm{U}_{\textrm{S},i}^H\bm{U}_{\textrm{S},i}\bm{\Lambda}_{\textrm{S},i}\bm{V}_{\textrm{S},i}^H{\bm{V}_{\textrm{S},i}}\textbf{w}_{\textrm{s}}^{(i)}\bm{s}_i
\\\nonumber
&\times\bm{s}_i^H{\textbf{w}_{\textrm{s}}^{(i)}}\bm{V}_{\textrm{S},i}^H\bm{V}_{\textrm{S},i}\bm{\Lambda}_{\textrm{S},i}\bm{U}_{\textrm{S},i}^H\bm{U}_{\textrm{S},i}\bm{\rho}^{\frac{1}{2}}),\nonumber\\
=&{P}_{\textrm{S},i}\parallel\bm{\rho}_i^{\frac{1}{2}}\bm{\Lambda}_{\textrm{S},i}{\textbf{w}_{\textrm{s}}^{(i)}}\parallel^2
\end{flalign}
Further, it is rewritten to be
\begin{flalign}\label{Eq:psEr2}
\mathcal {E}_{\textrm{R},i}
=\sum\nolimits_{n=1}^{N}{\rho_{i,n}}{\lambda^{(i)}_{\textrm{\tiny S},n}{P}_{\textrm{S},i}\textit{w}_{\textrm{s},n}^{(i)}},
\end{flalign}
where $\lambda^{(i)}_{\textrm{\tiny S},n}{P}_{\textrm{S},i}\textit{w}_{\textrm{s},n}^{(i)}$ can be treated as the harvested energy on the $n$-th spatial subchannel over the $i$-th subcarrier, i.e., the harvested energy over the $\ell$-th subchannel of the first hop. Therefore, the total energy harvested at ${\rm R}$ can be given by
\begin{flalign}\label{Eq:psEr2}
\mathscr{E}_{\textrm{R}}
&=\sum\nolimits_{i=1}^{K}\mathcal {E}_{\textrm{R},i}=\sum\nolimits_{i=1}^{K}\sum\nolimits_{n=1}^{N}{\rho_{i,n}}{\lambda^{(i)}_{\textrm{\tiny S},n}{P}_{\textrm{S},i}\textit{w}_{\textrm{s},n}^{(i)}},\nonumber\\
&=\sum\nolimits_{\ell=1}^{KN}{\rho_{\ell}}{\lambda_{\textrm{\tiny S},\ell}{\mathcal {P}}_{\textrm{S}}\omega_{\ell}},
\end{flalign}
where $\ell=(i-1)K+n$ and  $\omega_{\ell}\in [0,1]$ denote the power allocating factor at ${\rm S}$ over subchannel $\ell$.

In the meantime, the rest $(1-\rho_{i,n})$ power is split to the information receiver of ${\rm R}$ at the $n$-th spatial subchannel over the $i$-th subcarrier. Thus, the signal collected at the ``info receiver'' is
 \begin{equation}\label{Eq:YI}
\bm{y}_{\textrm{R},i}^{\textbf{(IF)}}=(\bm{I}-\bm{\rho})^{\frac{1}{2}}{\hat{\bm{\Lambda}}}_{\textrm{S},i}\hat{\bm{x}}_i
+\hat{\bm{U}}_{\textrm{S},i}^H\bm{z}_{\textrm{R},i}.
\end{equation}

As a result, the received signal at ${\rm D}$ can be given by
\begin{flalign}\label{Eq:psydnn}
&\bm{y}_{\textrm{D},j}=(\bm{I}-\bm{\rho})^{\frac{1}{2}}\hat{\bm{\Lambda}}_{\textrm{R},j}\bm{B}_{i}{\hat{\bm{\Lambda}}}_{\textrm{S},i}\hat{\bm{x}}_i
+\hat{\bm{\Lambda}}_{\textrm{R},j}\bm{B}_{i}\hat{\bm{U}}_{\textrm{S},i}^H\bm{z}_{\textrm{R},i}+\hat{\bm{U}}_{\textrm{R},j}^H\bm{z}_{\textrm{D},j}.\nonumber
\end{flalign}

In this paper, we assume that the energy harvested on the $\ell$-th subchannel over the ${\rm S}-{\rm R}$ link is only used for the information transmission on its paired subchannel, i.e., the $\ell'$-th subchannel of the ${\rm R}-{\rm D}$ link. Therefore, the available  transmit power for the $\ell'$-th subchannel at ${\rm R}$ is $p_{\textrm{\tiny R},\ell'}=\rho_{\ell}\lambda_{\textrm{\tiny S},\ell}{\mathcal {P}}_{\textrm{S}}\omega_{\ell}$. Since $p_{\textrm{\tiny S},\ell}\triangleq (1-\rho_{\ell})\mathcal{P}_{\textrm{S}}\omega_{\ell}$, $R_{\ell,\ell'}$ in (\ref{Eq:R}) then can be reexpressed as
\begin{flalign}
&R^{(\textrm{PSR})}_{\ell,\ell'}\\
&=\frac{\mathfrak{B}}{2K}\log\bigg(1+\frac{(1-\rho_{\ell})\rho_{\ell}\mathcal{P}_{\textrm{S}}^2\omega_{\ell}^2\lambda_{\textrm{\tiny S},\ell}\frac{\lambda_{\textrm{\tiny S},\ell}^2}{\sigma_{\textrm{\tiny R}}^2}\frac{\lambda_{\textrm{\tiny R},\ell'}}{\sigma_{\textrm{\tiny D}}^2}}
{1+(1-\rho_{\ell})\mathcal{P}_{\textrm{S}}\omega_{\ell}\frac{\lambda_{\textrm{\tiny S},\ell}}{\sigma_{\textrm{\tiny R}}^2}+\rho_{\ell}\lambda_{\textrm{\tiny S},\ell}{\mathcal {P}}_{\textrm{S}}\omega_{\ell}\frac{\lambda_{\textrm{\tiny R},\ell'}}{\sigma_{\textrm{\tiny D}}^2}}\bigg)\nonumber
\end{flalign}
for PSR. Consequently, the instantaneous achievable information rate of PSR can be given by
\begin{flalign}
C_{\textrm{PSR}}=\sum\limits_{\ell=1}^{KN}\sum\limits_{\ell'=1}^{KN} {\theta_{\ell,\ell'}{R^{\textrm{(PSR)}}_{\ell,\ell'}}},
\end{flalign}
where $\theta_{\ell,\ell'}\in\{0,1\}$ is the indicator for the subchannel-pairing, which has the same definition with that below (\ref{CTSR}).

\subsubsection{Optimization Problem Formulation for PSR}\label{Sec:OPTTSR}
To optimally design the proposed PSR, in this subsection, we formulate an optimization problem to jointly optimize
the power splitting factor vector $\bm{\rho}=\{\rho_{\ell}\}_{KN\times 1}$, the power allocating matrix $\bm{\omega}=\{\omega_{\ell}\}_{KN\times 1}$ at ${\rm S}$ and the subchannel pairing $\bm{\theta}=\{\theta_{\ell,\ell'}\}_{KN \times KN}$ over the two hops. The objective is also to find the optimal $\bm{\omega}^*$, $\bm{\theta}^*$ and $\bm{\rho}^*$ to maximize the E2E achievable information rate. Thus, the optimization problem can be expressed as
\begin{flalign}\label{Opt:PSR}
\mathop {\max }\limits_{\bm{\omega},\bm{\theta},\bm{\rho}}\,\, &C_{\textrm{PSR}}\\
\textrm{s.t.}\,\,&\sum\nolimits_{\ell=1}^{KN}{\omega_{\ell}}\leq 1,\,\,\theta_{\ell,\ell'}\in\{0,1\}\nonumber\\
&\sum\nolimits_{\ell=1}^{KN}{\theta_{\ell,\ell'}}=1, \forall \ell'; \sum\nolimits_{\ell'=1}^{KN}{\theta_{\ell,\ell'}}=1, \forall \ell\nonumber\\
&0\leq \rho_{\ell} \leq 1\nonumber, \forall \ell \in \{1,2,...,KN\},
\end{flalign}
where the constraint $\sum\nolimits_{\ell=1}^{KN}{\omega_{\ell}}\leq 1$ implies that the available transmit power at the source is constrained by $\mathcal{P}_{\textrm{S}}$. The constraints $\sum\nolimits_{\ell=1}^{KN}{\theta_{\ell,\ell'}}=1, \forall \ell'$ and $\sum\nolimits_{\ell'=1}^{KN}{\theta_{\ell,\ell'}}=1, \forall \ell$ mean that each subchannel of the first hop is allowed to be paired with only one subchannel of the second hop, vise verse. We shall discuss how to solve Problem (\ref{Opt:PSR}) in Section \ref{Sec:OPTDPSR}.

\section{Optimal Design of TSR}\label{Sec:OPTDTSR}
In this section, we shall investigate how to solve the optimization problem (\ref{Opt:TSR}) for TSR.
It can be observed that Problem (\ref{Opt:TSR}) is a combinatorial optimization problem with discrete variables $\theta_{\ell,\ell'}\in\{0,1\}$. Even if we remove the discrete variables $\theta_{\ell,\ell'}$, it is still a non-convex optimization problem, which cannot be solved by using conventional methods.  Thus, we solve it as follows.

Firstly, it can be seen that, in the first phase of TSR, only energy is transferred and accordingly only $\bm{\mathcal {X}}_{\textrm{S}}$ is required to be optimized, which means that $\bm{\mathcal {X}}_{\textrm{S}}$ is independent with other variables. Thus, it can be independently designed at first without loss of the global optimality.
Secondly, we found that the separation principle designed in \cite{R:Pairing} for joint channel pairing and power allocation optimization still holds in our system and $\bm{\theta}$ is also independent with $\alpha$ (We shall prove this in Section \ref{Sec:theta}). Therefore, $\bm{\theta}$ can also be optimized separately without jointly considering other variables. Based on these, we present a solution to Problem (\ref{Opt:TSR}) as shown in Algorithm \ref{alg:TSR}.  Then, we shall describe the detailed processing associated with each step of  Algorithm \ref{alg:TSR} in the successive subsections.

\begin{algorithm}[h]
  \caption{Optimization Framework for TSR}
  \label{alg:TSR}
  \begin{algorithmic}[1]
    \STATE Calculate the optimal $\bm{\mathcal {X}}_{\textrm{S}}^*$;
    \STATE Calculate the optimal $\bm{\theta}^*$;
    \STATE With the obtained $\bm{\mathcal {X}}_{\textrm{S}}^*$ and $\bm{\theta}^*$, find the optimal $\{\bm{\omega}^*, \bm{\varpi}^*\}$ and $\alpha^*$ to maximize $C_{\textrm{TSR}}$.
  \end{algorithmic}
\end{algorithm}

\subsection{Optimal $\bm{\mathcal {X}}_{\textrm{S}}^*$ for TSR}
From (\ref{Eq:PR}),  it can be seen that for a given $\alpha$, the larger $\sum\nolimits_{i=1}^{K}E_{\textrm{R},i}$, the higher $\mathcal{P}_{\textrm{R}}$, which means more available power for ${\rm R}$ to assist the information transmission from ${\rm S}$ to ${\rm D}$. This motivates that the objective of the optimal design of $\bm{\mathcal {X}}_{\textrm{S}}$ is to maximize $\sum\nolimits_{i=1}^{K}E_{\textrm{R},i}$. For a given $\alpha$, the optimization problem can be expressed by
\begin{flalign}\label{Opt:ProblemX}
\mathop{\max}\limits_{\bm{\mathcal {X}}_{\textrm{S}}}\,\,&\parallel\bm{H}_{\textrm{S},i}\bm{x}_i\parallel^2\\
&\sum\nolimits_{i=1}^{K}\textrm{tr}(\bm{X}_{i})\leq \mathcal{P}_{\textrm{S}}\nonumber\\
&\bm{X}_{i}\succeq 0\nonumber
\end{flalign}

 As mentioned in Section \ref{Sec:SectII}, by using SVD, $\bm{H}_{\textrm{S},i}=\bm{U}_{\textrm{S},i}\bm{\Lambda}_{\textrm{S},i}\bm{V}_{\textrm{S},i}^H$.
Let $\bm{v}_{s,1}^{(i)}$ be the first column of $\bm{V}_{\textrm{S},i}$.
With a given $\textrm{tr}(\bm{X}_{i})$, we can obtain the following Lemma 1 for solving Problem (\ref{Opt:ProblemX}).

\textbf{Lemma 1.} \textit{In TSR, for a given} $\textrm{tr}(\bm{X}_{i})$, \textit{the optimal} $\bm{X}_{i}^{\sharp}=\textrm{tr}(\bm{X}_{i})\bm{v}_{s,1}^{(i)}\bm{v}_{s,1}^{(i)H}$.
\begin{proof}
The proof of Lemma 1 can be found in Appendix \ref{App:L1} of this paper.
\end{proof}

Based on Lemma 1, we can further derive the following Lemma 2.

\textbf{Lemma 2.} \textit{In TSR, to achieve the maximum energy transfer, all power at ${\rm S}$ should be allocated to the subcarrier with the maximum} $\parallel\tilde{\bm{h}}^{(i)}_{\textrm{S},1}\parallel^2$ \textit{for all} $i\in \{1,2,...,K\}$.
\begin{proof}
The proof of Lemma 1 can be found in Appendix \ref{App:L2} of this paper.
\end{proof}

With Lemma 1 and Lemma 2, we can easily arrive at the following Theorem 1.

\textbf{Theorem 1.} \textit{The optimal solution of Problem (\ref{Opt:ProblemX}) is}
$\bm{\mathcal {X}}_{\textrm{S}}^*=\{\bm{X}_{i}^*\}$, \textit{where}
\begin{flalign}\label{Eq:OptX}
\bm{X}_{i}^*=\left\{ \begin{aligned}
 &\mathcal{P}_{\textrm{S}}\bm{v}_{s,1}^{(i)}\bm{v}_{s,1}^{(i)H},\,\,i=\textrm{arg} \max\limits_{b=1,...,K}{\parallel\tilde{\bm{h}}^{(b)}_{\textrm{S},1}\parallel^2},\\
 &0,\quad\quad\quad\quad\,\,\textit{otherwise},
 \end{aligned}\right.
\end{flalign}
\textit{and for a given $\alpha$, the optimal $\mathcal{P}_{\textrm{R}}$ is}
\begin{flalign}\label{Eq:OPr}
\mathcal{P}_{\textrm{R}}^{\sharp}=\frac{2\alpha }{1-\alpha}\mathcal{P}_{\textrm{S}}\parallel\tilde{\bm{h}}^{(c)}_{\textrm{S},1}\parallel^2,
\end{flalign}
\textit{where} $c=\textrm{arg} \max\limits_{b=1,...,K}{\parallel\tilde{\bm{h}}^{(b)}_{\textrm{S},1}\parallel^2}$ \textit{and} $i=1,...,K$.
\begin{proof}
By combining Lemma 1 and (\ref{Eq:OpttrX}), Theorem 1 can be easily proved.
\end{proof}

Theorem 1 indicates that the optimal power allocation for the energy transfer in TSR should be performed by projecting all the transmit energy contained in the space of the eigenvector with the largest eigenvalue of the channel matrix. Note that although the results in Lemma 1, Lemma 2 and Theorem 1 are derived for TSR in MIMO-OFDM relaying system, they just involve the energy transfer over the first hop, so these results also hold for multi-channel single-hop systems including MIMO and OFDM systems. Some similar results can also be seen in \cite{R:WR1}.

\subsection{Optimal $\bm{\theta}^*$ for TSR}\label{Sec:theta}
As described in (\ref{Eq:R}), the signal transmitted over the $\ell$-th subchannel of the first hop is allowed to be forwarded by ${\rm R}$ over the $\ell'$-th subcarrier of the second hop, which inspires subchannel pairing between the two hops. Substituting $\bm{\mathcal {X}}_{\textrm{S}}^*$ into Problem (\ref{Opt:TSR}), for a given $\alpha$, the optimization problem is rewritten to be
\begin{flalign}\label{Opt:TSR-II}
\mathop {\max }\limits_{\bm{\omega},\bm{\varpi},\bm{\theta}}\,\, &\frac{\mathfrak{B}}{2K}\frac{1-\alpha}{2}\sum\limits_{\ell=1}^{KN}{\theta_{\ell,\ell'}\log\bigg(1+\tfrac{\mathcal{P}_{\textrm{S}}\omega_{\ell}\mathcal{P}_{\textrm{R}}^{\sharp}\varpi_{\ell}\frac{\lambda_{\textrm{\tiny S},\ell}}{\sigma_{\textrm{\tiny R}}^2}\frac{\lambda_{\textrm{\tiny R},\ell'}}{\sigma_{\textrm{\tiny D}}^2}}
{1+\mathcal{P}_{\textrm{S}}\omega_{\ell}\frac{\lambda_{\textrm{\tiny S},\ell}}{\sigma_{\textrm{\tiny R}}^2}+\mathcal{P}_{\textrm{R}}^{\sharp}\varpi_{\ell}\frac{\lambda_{\textrm{\tiny R},\ell'}}{\sigma_{\textrm{\tiny D}}^2}}\bigg)}\nonumber\\
\textrm{s.t.}\,\,&\sum\nolimits_{\ell=1}^{KN}{\omega_{\ell}}\leq 1,\,\,\sum\nolimits_{\ell'=1}^{KN}{\varpi_{\ell}}\leq 1\nonumber\\
&\sum\nolimits_{\ell=1}^{KN}{\theta_{\ell,\ell'}}=1, \forall \ell', \,\,\sum\nolimits_{\ell'=1}^{KN}{\theta_{\ell,\ell'}}=1, \forall \ell
\end{flalign}

Problem (\ref{Opt:TSR-II}) can be regarded as a joint power allocation and subchannel pairing (JPASP) problem\cite{R:Pairingk,R:Pairing}.
In \cite{R:Pairing}, it was proved that the JPASP problem can be solved in a separated manner without losing the global optimality, where the channel pairing can be determined with sorted channel gain. That is, the subchannel with the $i$-th largest channel gain over the first hop should be paired with the subchannel with the $i$-th largest channel gain over the second hop. Therefore, by optimally allocating the power over the paired subchannels, the obtained result is the same with that obtained by jointly optimized power allocation and subchannel pairing. Based on this, we can derive the following Lemma 3.

\textbf{Lemma 3.} \textit{In TSR, the optimal} $\bm{\theta}^*$ \textit{satisfies that}
\begin{flalign}\label{Eq:Theta}
{\theta}_{\ell,\ell'}^*=\left\{ \begin{aligned}
 &1,\,\, \textit{if}\,\,\textit{Order}(\lambda_{\textrm{S},\ell}) = \textit{Order}(\lambda_{\textrm{R},\ell'}),\\
 &0,\,\,\textit{otherwise}.
 \end{aligned}\right.
 \end{flalign}
\textit{where }$\textit{Order}(\lambda_{u,\ell})$ \textit{represents the rank position of} $\lambda_{u,\ell}$ \textit{among $\lambda_{u,i}$ for all $i=1,...,KN$ with a descending sorting order at node $u\in\{\textrm{S},\textrm{R}\}$}.
\begin{proof}
With the decoupling policy presented in \cite{R:Pairing}, we can easily prove that the optimal $\theta$ for Problem (\ref{Opt:TSR-II}) is (\ref{Eq:Theta}). Moreover, it can be inferred that the change of $\alpha$ only affects the available power $\mathcal{P}_{\textrm{R}}^{\sharp}$ and the time duration for information transmission, which does not affect the channel gain of all subchannels, so the result in (\ref{Eq:Theta}) is also the global optimal solution for the original Problem (\ref{Opt:TSR}).
\end{proof}

\subsection{Joint optimal  $\bm{\omega}^*,\bm{\varpi}^*$ and $\alpha^*$}
With the optimal subchannel pairing $\bm{\theta}^*$ obtained in Lemma 3 and the optimal design of $\bm{\mathcal {X}}_{\textrm{S}}^*$ described in Theorem 1, $C_{\textrm{TSR}}$ can be considered as the sum of rate over $KN$ independent E2E paths. However, it is still neither joint convex nor joint concave w.r.t. $\bm{\omega}$, $\bm{\varpi}$ and $\alpha$. Therefore, we adopt the following method to solve it. Firstly, for a given $\alpha$, we find the optimal $\bm{\omega}^*$,$\bm{\varpi}^*$. Then, we consider  $\bm{\omega}^*$ and $\bm{\varpi}^*$ as two functions of $\alpha$, $\bm{\omega}^*(\alpha)$ and $\bm{\varpi}^*(\alpha)$.  By substituting $\bm{\omega}^*(\alpha)$ and $\bm{\varpi}^*(\alpha)$ for $\bm{\omega}$ and $\bm{\varpi}$, respectively, we then calculate the optimal $\alpha^*$. The detailed operations are described in the following subsections.

\subsubsection{Optimal  $\{\bm{\omega}^*,\bm{\varpi}^*\}$ for a given $\alpha$}
For a given $\alpha$, the optimization problem (\ref{Opt:TSR}) can be rewritten to be
\begin{flalign}\label{Opt:PowTSR}
\mathop {\max }\limits_{\bm{\omega},\bm{\varpi}}\,\, &\frac{1-\alpha}{2}\sum\nolimits_{\ell=1}^{KN}{R_{\ell}}\nonumber\\
\textrm{s.t.}\,\,&\sum\nolimits_{\ell=1}^{KN}{\omega_{\ell}}\leq 1,\,\,\sum\nolimits_{\ell=1}^{KN}{\varpi_{\ell}}\leq 1,
\end{flalign}
where $R_{\ell}=\frac{\mathfrak{B}}{2K}\log\bigg(1+\tfrac{\mathcal{P}_{\textrm{S}}\omega_{\ell}{P}_{\textrm{R}}^{\sharp}\varpi_{\ell}\tfrac{\lambda_{\textrm{\tiny S},\ell}}{\sigma_{\textrm{\tiny R}}^2}\tfrac{\lambda_{\textrm{\tiny R},\ell}}{\sigma_{\textrm{\tiny D}}^2}}
{1+\mathcal{P}_{\textrm{S}}\omega_{\ell}\tfrac{\lambda_{\textrm{\tiny S},\ell}}{\sigma_{\textrm{\tiny R}}^2}+\mathcal{P}_{\textrm{R}}^{\sharp}\varpi_{\ell}\tfrac{\lambda_{\textrm{\tiny R},\ell}}{\sigma_{\textrm{\tiny D}}^2}}\bigg)$.

It can be proved that the objective function  of Problem (\ref{Opt:PowTSR}) is not jointly concave in
$\omega_{\ell}$ and $\varpi_{\ell}$. Therefore, it is not possible to obtain the global optimal solution analytically. However, we find that if $\omega_{\ell}$ is fixed, the objective function of Problem (\ref{Opt:PowTSR}) is concave in $\varpi_{\ell}$, vice versa.
Therefore, Karush-Kuhn-Tucker (KKT) \cite{R:CVX} can be applied to derive the optimal solutions. As a result, we arrive at Lemma 4.

\textbf{Lemma 4.} \textit{In TSR, for a given $\alpha$ and $\varpi_{\ell}$ $(\ell=\{1,2...,KN\})$, the optimal $\omega_{\ell}$ is}
\begin{flalign}\label{Eq:OMe}
\omega_{\ell}^{\sharp}=\frac{\sigma_{\textrm{\tiny R}}^2}{\mathcal{P}_{\textrm{S}}\cdot\lambda_{\textrm{\tiny S},\ell}}\Bigg[\frac{\mathcal{P}_{\textrm{R}}^{\sharp}\varpi_{\ell}\lambda_{\textrm{\tiny R},\ell}}{2\sigma_{\textrm{\tiny D}}^2}
\Bigg(\sqrt{1+\frac{4\lambda_{\textrm{\tiny S},\ell}\sigma_{\textrm{\tiny D}}^2}{\sigma_{\textrm{\tiny R}}^2\lambda_{\textrm{\tiny R},\ell}\mathcal{P}_{\textrm{R}}^{\sharp}\varpi_{\ell}\mu}}-1\Bigg)-1\Bigg]^+,
\end{flalign}
\textit{while for given $\varpi_{\ell}$ $(\omega=\{1,2...,KN\})$, the optimal $\varpi_{\ell}$ is}
\begin{flalign}\label{Eq:Var}
\varpi_{\ell}^{\sharp}=\frac{\sigma_{\textrm{\tiny D}}^2}{\mathcal{P}_{\textrm{R}}^{\sharp}\cdot\lambda_{\textrm{\tiny R},\ell}}\Bigg[\frac{\mathcal {P}_{\textrm{S}}\omega_{\ell}\lambda_{\textrm{\tiny S},\ell}}{2\sigma_{\textrm{\tiny R}}^2}
\Bigg(\sqrt{1+\frac{4\lambda_{\textrm{\tiny R},\ell}\sigma_{\textrm{\tiny R}}^2}{\sigma_{\textrm{\tiny D}}^2\lambda_{\textrm{\tiny S},\ell}\mathcal {P}_{\textrm{S}}\omega_{\ell}\nu}}-1\Bigg)-1\Bigg]^+,
\end{flalign}
\textit{where $[x]^+=\max\{0,x\}$. $\mu$ and $\nu$ are defined as non-negative Lagrange multipliers, which have to be chosen such that $\sum\nolimits_{\ell=1}^{KN}{\omega_{\ell}^{\sharp}}\leq 1,$ and $\sum\nolimits_{\ell=1}^{KN}{\varpi_{\ell}^{\sharp}}\leq 1$, respectively.}

According to Lemma 4, we design such an iterative method, as shown in Algorithm \ref{alg:ww}, to get the near optimal solution for Problem (\ref{Opt:PowTSR}) with a given bias error $\epsilon$.
\begin{algorithm}[h]
\caption{Finding the local or global optimal $\{\bm{\omega}^*,\bm{\varpi}^*\}$}\label{alg:ww}
\begin{algorithmic}[1]
\FOR{each $\ell \in [1,KN]$}
\STATE Initialize $\omega_{\ell}=\frac{1}{KN}$;\
\STATE Calculate $\varpi_{\ell}$ in terms of (\ref{Eq:Var});\
\ENDFOR
\STATE Initialize $C_{pre}=0$;\
\STATE Calculate
 $C_{cur}=\frac{1-\alpha}{2}\sum\nolimits_{\ell=1}^{KN}{R_{\ell}}$;\
\label{code:recentStart}
\WHILE {$|C_{cur}-C_{pre}|>\epsilon$}
\STATE Update $C_{pre}=C_{cur}$;\
\FOR{each $\ell \in [1,KN]$}
\STATE Update $\omega_{\ell}$ in terms of (\ref{Eq:OMe});\
\STATE Update $\varpi_{\ell}$ in terms of (\ref{Eq:Var});\
\STATE Update $C_{cur}=\frac{1-\alpha}{2}\sum\nolimits_{\ell=1}^{KN}{R_{\ell}}$;\
\ENDFOR
\ENDWHILE
\STATE Return $\{\bm{\omega},\bm{\varpi}\}$.
\label{code:recentEnd}
\end{algorithmic}
\end{algorithm}

Since $C_{cur}$ in Algorithm \ref{alg:ww} is concave w.r.t in $\omega_{\ell}$ and $\varpi_{\ell}$,  each round of iteration Algorithm \ref{alg:ww} can improve $C_{cur}$. As $\sum\nolimits_{\ell=1}^{KN}{\omega_{\ell}}\leq 1$ and $\sum\nolimits_{\ell=1}^{KN}{\varpi_{\ell}}\leq 1$, $C_{cur}$ cannot be increased without limit. This implies the \textit{convergence} of Algorithm \ref{alg:ww}.
Moreover, it also can be observed that Algorithm \ref{alg:ww} depends on the initialization of $\omega_{\ell}$, although we adopt the equal weight of $\omega_{\ell}$ for it, Algorithm \ref{alg:ww} cannot always guarantee the global optimality. Therefore,  we shall also investigate an asymptotically global optimal solution of $\{\bm{\omega}^*,\bm{\varpi}\}^*$ as follows for high SNR case.

At high SNR region, we can approximate $R_\ell$ as
\begin{flalign}
R_\ell&=\frac{\mathfrak{B}}{2K}\log\bigg(1+\frac{\mathcal{P}_{\textrm{S}}\omega_{\ell}\mathcal{P}_{\textrm{R}}^{\sharp}\varpi_{\ell}\frac{\lambda_{\textrm{\tiny S},\ell}}{\sigma_{\textrm{\tiny R}}^2}\frac{\lambda_{\textrm{\tiny R},\ell}}{\sigma_{\textrm{\tiny D}}^2}}
{1+\mathcal{P}_{\textrm{S}}\omega_{\ell}\frac{\lambda_{\textrm{\tiny S},\ell}}{\sigma_{\textrm{\tiny R}}^2}+\mathcal{P}_{\textrm{R}}^{\sharp}\varpi_{\ell}\frac{\lambda_{\textrm{\tiny R},\ell}}{\sigma_{\textrm{\tiny D}}^2}}\bigg)
\nonumber\\
&\simeq \frac{\mathfrak{B}}{2K}\log\bigg(1+\frac{\mathcal{P}_{\textrm{S}}\omega_{\ell}\mathcal{P}_{\textrm{R}}^{\sharp}\varpi_{\ell}\frac{\lambda_{\textrm{\tiny S},\ell}}{\sigma_{\textrm{\tiny R}}^2}\frac{\lambda_{\textrm{\tiny R},\ell}}{\sigma_{\textrm{\tiny D}}^2}}
{\mathcal{P}_{\textrm{S}}\omega_{\ell}\frac{\lambda_{\textrm{\tiny S},\ell}}{\sigma_{\textrm{\tiny R}}^2}+\mathcal{P}_{\textrm{R}}^{\sharp}\varpi_{\ell}\frac{\lambda_{\textrm{\tiny R},\ell}}{\sigma_{\textrm{\tiny D}}^2}}\bigg)
\end{flalign}

Such an approximation leads to the Hessian Matrix  $\big[\frac{\partial^2 R_\ell}{\partial \omega_{\ell}}, \frac{\partial^2 R_\ell}{\partial \varpi_{\ell}}\big]^T \preceq \textbf{0}$, which indicates a jointly concave $R_\ell$ in $\omega_{\ell}$ and $\varpi_{\ell}$. In this case, we can give the optimal solution as described in Lemma 5 by using KKT conditions.

\textbf{Lemma 5.} \textit{In TSR, at high SNR regime, for a given $\alpha$, the optimal $\omega_{\ell}^*$ and $\omega_{\ell}^*$ can be approximated by}
\begin{flalign}\label{Eq:OMe}
\omega_{\ell}^\star=\frac{1}{{\mathcal{P}_{\textrm{S}}\Big(1+\sqrt{\tfrac{\lambda_{\textrm{\tiny S},\ell}\sigma_{\textrm{\tiny D}}^2}{\sigma_{\textrm{\tiny R}}^2\lambda_{\textrm{\tiny R},\ell}}}}\Big)}\Bigg[\frac{1}{\mu}-\tfrac{\big(\sqrt{\tfrac{\lambda_{\textrm{\tiny S},\ell}}{\sigma_{\textrm{\tiny R}}^2}}+\sqrt{\tfrac{\lambda_{\textrm{\tiny R},\ell}}{\sigma_{\textrm{\tiny D}}^2}}\big)^2}{\tfrac{\lambda_{\textrm{\tiny S},\ell}}{\sigma_{\textrm{\tiny R}}^2}\tfrac{\lambda_{\textrm{\tiny R},\ell}}{\sigma_{\textrm{\tiny D}}^2}}\Bigg]^+
\end{flalign}
\textit{and}
\begin{flalign}\label{Eq:OMe}
\varpi_{\ell}^\star=\frac{1}{{\mathcal{P}^{\sharp}_{\textrm{R}}\Big(1+\sqrt{\tfrac{\lambda_{\textrm{\tiny R},\ell}\sigma_{\textrm{\tiny R}}^2}{\sigma_{\textrm{\tiny D}}^2\lambda_{\textrm{\tiny S},\ell}}}}\Big)}\Bigg[\frac{1}{\nu}-\frac{\big(\sqrt{\tfrac{\lambda_{\textrm{\tiny S},\ell}}{\sigma_{\textrm{\tiny R}}^2}}+\sqrt{\tfrac{\lambda_{\textrm{\tiny R},\ell}}{\sigma_{\textrm{\tiny D}}^2}}\big)^2}{\tfrac{\lambda_{\textrm{\tiny S},\ell}}{\sigma_{\textrm{\tiny R}}^2}\tfrac{\lambda_{\textrm{\tiny R},\ell}}{\sigma_{\textrm{\tiny D}}^2}}\Bigg]^+,
\end{flalign}
\textit{respectively, where $\mu$ and $\nu$ are two positive Lagrangian parameters, which have to be chosen such that $\sum\nolimits_{\ell=1}^{KN}{\omega_{\ell}^\star}\leq 1,$ and $\sum\nolimits_{\ell=1}^{KN}{\varpi_{\ell}^\star}\leq 1$.}

\subsubsection{Optimal $\alpha^*$}
From Lemma 4 and Lemma 5, we can see that each of $\bm{\omega}^*$, $\bm{\varpi}^*$, $\bm{\omega}^\star$ and $\bm{\varpi}^\star$ has close relationship with $\alpha$. Thus, each of them can be regarded as a function w.r.t $\alpha$.
Substituting $\{\bm{\omega}^*,\bm{\varpi}^*\}$ into (\ref{Opt:PowTSR}), we have that
\begin{flalign}\label{Opt:Ca}
&C_{\textrm{TSR}}(\alpha)=\nonumber\\
&\frac{\mathfrak{B}}{2K}\tfrac{1-\alpha}{2}\sum\limits_{\ell=1}^{KN}{\log\bigg(1+\tfrac{\mathcal{P}_{\textrm{S}}\omega_{\ell}^*(\alpha)\mathcal{P}_{\textrm{R}}^{\sharp}(\alpha)\varpi_{\ell}^*(\alpha)\frac{\lambda_{\textrm{\tiny S},\ell}}{\sigma_{\textrm{\tiny R}}^2}\tfrac{\lambda_{\textrm{\tiny R},\ell}}{\sigma_{\textrm{\tiny D}}^2}}
{1+\mathcal{P}_{\textrm{S}}\omega_{\ell}^*(\alpha)\frac{\lambda_{\textrm{\tiny S},\ell}}{\sigma_{\textrm{\tiny R}}^2}+\mathcal{P}_{\textrm{R}}^{\sharp}(\alpha)\varpi_{\ell}^*(\alpha)\frac{\lambda_{\textrm{\tiny R},\ell}}{\sigma_{\textrm{\tiny D}}^2}}\bigg)}.
\end{flalign}

Let
\begin{flalign}\label{Opt:G}
\left\{ \begin{aligned}
&\mathcal{G}(\alpha)=\sum\limits_{\ell=1}^{KN}{\log\bigg(1+\tfrac{\mathcal{P}_{\textrm{S}}\omega_{\ell}^*(\alpha)\mathcal{P}_{\textrm{R}}^{\sharp}(\alpha)\varpi_{\ell}^*(\alpha)\frac{\lambda_{\textrm{\tiny S},\ell}}{\sigma_{\textrm{\tiny R}}^2}\frac{\lambda_{\textrm{\tiny R},\ell}}{\sigma_{\textrm{\tiny D}}^2}}
{1+\mathcal{P}_{\textrm{S}}\omega_{\ell}^*(\alpha)\frac{\lambda_{\textrm{\tiny S},\ell}}{\sigma_{\textrm{\tiny R}}^2}+\mathcal{P}_{\textrm{R}}^{\sharp}(\alpha)\varpi_{\ell}^*(\alpha)\frac{\lambda_{\textrm{\tiny R},\ell}}{\sigma_{\textrm{\tiny D}}^2}}\bigg)}\\
&\mathcal{G'}(\alpha,\alpha')=\sum\limits_{\ell=1}^{KN}{\log\bigg(1+\tfrac{\mathcal{P}_{\textrm{S}}\omega_{\ell}^*(\alpha)\mathcal{P}_{\textrm{R}}^{\sharp}(\alpha')\varpi_{\ell}^*(\alpha)\frac{\lambda_{\textrm{\tiny S},\ell}}{\sigma_{\textrm{\tiny R}}^2}\frac{\lambda_{\textrm{\tiny R},\ell}}{\sigma_{\textrm{\tiny D}}^2}}
{1+\mathcal{P}_{\textrm{S}}\omega_{\ell}^*(\alpha)\frac{\lambda_{\textrm{\tiny S},\ell}}{\sigma_{\textrm{\tiny R}}^2}+\mathcal{P}_{\textrm{R}}^{\sharp}(\alpha')\varpi_{\ell}^*(\alpha)\frac{\lambda_{\textrm{\tiny R},\ell}}{\sigma_{\textrm{\tiny D}}^2}}\bigg)}.
\end{aligned}\right.
\end{flalign}
We can derive the following Lemma 6.

\textbf{Lemma 6.} \textit{$\mathcal{G}(\alpha)$ is a monotonically increasing function w.r.t variable $\alpha\in [0,1]$}.
\begin{proof}
Suppose the two variables $\alpha$ and $\alpha'$, $0\leq\alpha<\alpha'\leq 1$, then we have $\mathcal{G}(\alpha)<\mathcal{G}(\alpha')$.
As $\alpha<\alpha'$, we can deduce that $\mathcal{P}_{\textrm{R}}^{\sharp}(\alpha)<\mathcal{P}_{\textrm{R}}^{\sharp}(\alpha')$, which implies that
$\mathcal{G}(\alpha)<\mathcal{G'}(\alpha,\alpha')$. Moreover, for given $\alpha'$, $\omega_{\ell}^*(\alpha')$ and $\varpi_{\ell}^*(\alpha')$ are optimized to increase $\mathcal{G}(\alpha')$, so it can be inferred that $\mathcal{G}(\alpha')\geq \mathcal{G'}(\alpha,\alpha')$.
\end{proof}

With the definition of $\mathcal{G}(\alpha)$, one can rewritten $C_{\textrm{TSR}}(\alpha)$ as $C_{\textrm{TSR}}(\alpha)=\frac{\mathfrak{B}}{2K}\frac{1-\alpha}{2}\mathcal{G}(\alpha)$, which is a product of a monotonic decreasing function and a monotonic increasing function. Besides, it can be easily observed that
$C_{\textrm{TSR}}(0)=C_{\textrm{TSR}}(1)=0$ and $C_{\textrm{TSR}}(\alpha)>0$ for $\alpha\in (0,1])$. Thus, there exists a maximum value of $C_{\textrm{TSR}}(\alpha)$ within the interval $\alpha\in (0,1]$.
Nevertheless, it is still very difficult to analytically discuss the convexity of $C_{\textrm{TSR}}(\alpha)$ due to the implicit expression of $\omega_{\ell}^*(\alpha)$ and $\varpi_{\ell}^*(\alpha)$. As our goal is to explore the potential capacity of the TSR, we adopt a numerical method to search the maximum $C_{\textrm{TSR}}(\alpha^*)$ over $\alpha\in (0,1])$ with an updating step $\Delta \alpha$, where the computational complexity is about $\mathcal{O}(\frac{1}{\Delta \alpha})$. Note that in our simulations, we found that $C_{\textrm{TSR}}(\alpha)$ is always a firstly increasing and then decreasing function of $\alpha$, which indicates that $C_{\textrm{TSR}}(\alpha)$ has only one peak (or maximum) within $\alpha\in [0,1]$. Therefore, some conventional algorithms with fast convergence such as hill climbing algorithm \cite{R:CH} also can be adopted to find $\alpha^*$.

\section{Optimal Design of PSR Protocol}\label{Sec:OPTDPSR}
In this Section, we shall investigate how to solve the optimization problem (\ref{Opt:PSR}) for PSR.  It also can be seen that the Problem (\ref{Opt:PSR}) is with discrete variables $\theta_{\ell,\ell'}\in\{0,1\}$, which leads to a combinatorial optimization problem with high computational complexity. Thus, it cannot be easily solved by using conventional methods. In this subsection, we shall solve it as follows.

Firstly, we found that the optimal subchannel pairing strategy for TSR also holds for PSR, so that the optimal subchannel pairing can also be performed separately, which greatly reduces the complexity. Moreover, for a given $\omega_{\ell}$, we derive the explicit expression of optimal $\rho_{\ell}^*$, which can be regarded as a function w.r.t $\omega_{\ell}$, i.e., $\rho_{\ell}^*(\omega_{\ell})$. Thus, we replace the variable $\rho_{\ell}$ with $\rho_{\ell}^*(\omega_{\ell})$ to reduce the number of variables required to be optimized and then derive the optimal $\omega_{\ell}^*$.

For clarity, we present the optimization framework of PSR  as shown in Algorithm \ref{Alg:PSR} at first. Then we shall explain the detailed operation of each step of the framework in the successive subsections.

\begin{algorithm}[h]
  \caption{Optimization Framework for PSR}
  \label{Alg:PSR}
  \begin{algorithmic}[1]
    \STATE Calculate the optimal $\bm{\theta}^*$;
    \STATE Calculate the optimal $\bm{\rho}^*$ for a given $\bm{\omega}$;
    \STATE Calculate the optimal $\bm{\omega}^*$ with the obtained $\bm{\rho}^*$.
  \end{algorithmic}
\end{algorithm}

\subsection{Optimal $\bm{\theta}^*$ for PSR}
We find that Lemma 3 still holds for PSR. The reason is that, with the decoupling policy presented in \cite{R:Pairing}, the optimal subchannel pairing for PSR can also be obtained with the sorted channel gain of the subchannels. Moreover, since both $\bm{\rho}$ and $\bm{\omega}$ do not affect the result of the  sorted channel gain of the subchannels over the two hops, the optimal $\bm{\theta}^*$ can be independently obtained without considering the other variables. Thus, the optimal $\bm{\theta}^*$ for PSR also can be determined according to Lemma 3.

\subsection{Optimal $\bm{\rho}^*$ for a given $\bm{\omega}$}
Substituting $\bm{\theta}^*$ into problem (\ref{Opt:PSR}), it can be rewritten as
\begin{flalign}\label{Opt:PSR2}
\mathop {\max }\limits_{\bm{\omega},\bm{\rho}}\,\, &\frac{\mathfrak{B}}{2K}\sum\limits_{\ell=1}^{KN}{\log\bigg(1+\tfrac{(1-\rho_{\ell})\rho_{\ell}\mathcal{P}_{\textrm{S}}^2\omega_{\ell}^2\frac{\lambda_{\textrm{\tiny S},\ell}^2}{\sigma_{\textrm{\tiny R}}^2}\frac{\lambda_{\textrm{\tiny R},\ell'}}{\sigma_{\textrm{\tiny D}}^2}}
{1+(1-\rho_{\ell})\mathcal{P}_{\textrm{S}}\omega_{\ell}\frac{\lambda_{\textrm{\tiny S},\ell}}{\sigma_{\textrm{\tiny R}}^2}+\rho_{\ell}\lambda_{\textrm{\tiny S},\ell}{\mathcal {P}}_{\textrm{S}}\omega_{\ell}\frac{\lambda_{\textrm{\tiny R},\ell'}}{\sigma_{\textrm{\tiny D}}^2}}\bigg)}\nonumber\\
\textrm{s.t.}\,\,&\sum\nolimits_{\ell=1}^{KN}{\omega_{\ell}}\leq 1,\nonumber\\
&0\leq \rho_{\ell} \leq 1, \forall \ell \in \{1,2,..., KN\}
\end{flalign}

Since Problem (\ref{Opt:PSR2}) is also neither joint concave nor convex w.r.t $\bm{\omega}$ and $\bm{\rho}$, we shall discuss it with a given $\bm{\omega}$ at first.

\textbf{Theorem 2.} \textit{Given a power allocation weight vector $\bm{\omega}$ at source ${\rm S}$, the conditional optimal power splitting vector $\bm{\rho}^{\sharp}$ satisfies that}
\begin{flalign}\label{Eq:optrho}
\rho_{\ell}^{\sharp}=\left\{
\begin{aligned}
&\frac{A_{\ell}\omega_{\ell} + 1 - \sqrt{(A_{\ell}\omega_{\ell} +1)( Q_{\ell}\omega_{\ell} + 1)}}{A_{\ell}\omega_{\ell} - Q_{\ell}\omega_{\ell}},\,\, A_{\ell} \neq Q_{\ell} \\
&\frac{1}{2},\quad\quad\quad\quad\quad\quad\quad\quad\quad\quad\quad\quad\quad\quad\quad\,\, A_{\ell} = Q_{\ell},
\end{aligned}
\right.
\end{flalign}
\textit{where} $A_{\ell}=\mathcal{P}_{\textrm{S}}\frac{\lambda_{\textrm{\tiny S},\ell}}{\sigma_{\textrm{\tiny R}}^2}$ \textit{and} $Q_{\ell}={\mathcal {P}}_{\textrm{S}}\frac{\lambda_{\textrm{\tiny S},\ell}\lambda_{\textrm{\tiny R},\ell'}}{\sigma_{\textrm{\tiny D}}^2}$.
\begin{proof}
The proof of Theorem 2 can be found in Appendix \ref{App:T2}.
\end{proof}

For a special case, where the system is with a single carrier and each node is with only one antenna, we can easily deduce the following corollary 1 from Theorem 2.

\textbf{\textrm{Corollary 1.}} \textit{For a single-carrier single-antenna two-hop non-regenerative PSR relaying system, if the channel gain-to noise ratio (CNR) of the two hops satisfies that
$\gamma_{\textrm{\tiny S},\textrm{\tiny R}}=\gamma_{\textrm{\tiny R},\textrm{\tiny D}}$, the optimal power splitting ratio is $\rho^*=0.5$, where $\gamma_{u,v}=\frac{h_{u,v}}{\sigma_{v}^2}$}.

\subsection{Optimal $\bm{\omega}^*$ for PSR}
By substituting (\ref{Eq:optrho}) into Problem (\ref{Opt:PSR2}), we then obtain that
\begin{flalign}\label{Opt:PSR3}
\mathop {\max }\limits_{\bm{\omega}}\,\, &\frac{\mathfrak{B}}{2K}\sum\limits_{\ell=1}^{KN}{\log\bigg(1+\tfrac{(1-\rho_{\ell}^{\sharp})\rho_{\ell}^{\sharp}\mathcal{P}_{\textrm{S}}^2\omega_{\ell}^2\frac{\lambda_{\textrm{\tiny S},\ell}^2}{\sigma_{\textrm{\tiny R}}^2}\frac{\lambda_{\textrm{\tiny R},\ell'}}{\sigma_{\textrm{\tiny D}}^2}}
{1+(1-\rho_{\ell}^{\sharp})\mathcal{P}_{\textrm{S}}\omega_{\ell}\frac{\lambda_{\textrm{\tiny S},\ell}}{\sigma_{\textrm{\tiny R}}^2}+\rho_{\ell}^{\sharp}\lambda_{\textrm{\tiny S},\ell}{\mathcal {P}}_{\textrm{S}}\omega_{\ell}\frac{\lambda_{\textrm{\tiny R},\ell'}}{\sigma_{\textrm{\tiny D}}^2}}\bigg)}\nonumber\\
\textrm{s.t.}\,\,&\sum\nolimits_{\ell=1}^{KN}{\omega_{\ell}}\leq 1.
\end{flalign}

From (\ref{Eq:optrho}), although $\rho_{\ell}^{\sharp}$ can be regarded as a function of $\omega_{\ell}$, it is still not a simple expression, which makes problem (\ref{Opt:PSR3}) too difficult to be solved by using conventional methods. Thus, we firstly discuss analytically and then design efficient algorithm to solve it.

\textbf{Lemma 8.} \textit{Let
\begin{equation}\nonumber
r_{\ell}(\omega_{\ell})=\log\bigg(1+\frac{(1-\rho_{\ell}(\omega_{\ell}))\rho_{\ell}(\omega_{\ell})\omega_{\ell}^2A_{\ell}Q_{\ell}}
{1+(1-\rho_{\ell})(\omega_{\ell})\omega_{\ell}A_{\ell}+\rho_{\ell}(\omega_{\ell})\omega_{\ell}Q_{\ell}}\bigg),
\end{equation}
 where $\rho_{\ell}(\omega_{\ell})$ is a function w.r.t. $\omega_{\ell}$, as shown in (\ref{Eq:optrho}).  $r_{\ell}$ is a monotonically increasing function of $\omega_{\ell}$}.
\begin{proof}
We can derive $\frac{\partial r_{\ell}}{\partial \omega_{\ell}}$ with the consideration of cases $A_{\ell}=Q_{\ell}$ and $A_{\ell}\neq Q_{\ell}$, respectively, as follows.
\begin{flalign}\label{Eq:Diffr}
\frac{\partial r_{\ell}}{\partial \omega_{\ell}}=
\left\{ \begin{aligned}
&\tfrac{\mathcal{Z}_{\ell}(\mathcal {C}_{\ell}-\mathcal {G}_{\ell})}{\mathcal {B}_{\ell}},\quad\quad\quad\quad\quad\quad\quad\quad \quad\,\,  \textrm{for}\,\,A_{\ell}\neq Q_{\ell}\\
&\tfrac{A_{\ell}Q_{\ell}\omega_{\ell}(A_{\ell}\omega_{\ell} + Q_{\ell}\omega_{\ell} + 4)}{(A_{\ell}\omega_{\ell} + 2)(Q_{\ell}\omega_{\ell} + 2)(A_{\ell}\omega_{\ell} + Q_{\ell}\omega_{\ell} + 2)},\,\,\,\textrm{for}\,\, A_{\ell}= Q_{\ell}
 \end{aligned}\right.
\end{flalign}
where
\begin{small}
\begin{flalign}\label{Eq:Upsilon}
\left\{ \begin{aligned}
 &\mathcal{Z}_{\ell}=AQ(A - Q)^2(A_{\ell}\omega_{\ell} + 1)(Q_{\ell}\omega_{\ell} + 1)(A_{\ell} + Q_{\ell} + AQ_{\ell}\omega_{\ell})\\
 &\mathcal {C}_{\ell}=(A_{\ell}\omega_{\ell} + Q_{\ell}\omega_{\ell} + AQ_{\ell}\omega_{\ell}^2 + 1)^\frac{3}{2}\\
 &\mathcal {G}_{\ell}=(Q_{\ell}\omega_{\ell} + 1)^2(A_{\ell}\omega_{\ell} + 1)^2\\
 &\mathcal {B}_{\ell}=\\
 &-(Q_{\ell}\omega_{\ell} + 1)^3(A_{\ell}\omega_{\ell} + 1)^3(A_{\ell} - Q_{\ell})^2(A_{\ell} + Q_{\ell} + A_{\ell}Q_{\ell}\omega_{\ell})^2.
 \end{aligned}\right.
 \end{flalign}
 \end{small}
 With some algebraic manipulation, Lemma 8 can be easily proved.
 \end{proof}

\textbf{Lemma 9.} \textit{Let
\begin{equation}\nonumber
r_{\ell}(\omega_{\ell})=\log\bigg(1+\frac{(1-\rho_{\ell}(\omega_{\ell}))\rho_{\ell}(\omega_{\ell})\omega_{\ell}^2A_{\ell}Q_{\ell}}
{1+(1-\rho_{\ell})(\omega_{\ell})\omega_{\ell}A_{\ell}+\rho_{\ell}(\omega_{\ell})\omega_{\ell}Q_{\ell}}\bigg).
\end{equation}
It can be observed that $r_{\ell}(0)=0$}.

Although we have obtained some disciplines on Problem (\ref{Opt:PSR3}), it is still very difficult to
prove the convexity of $\sum\nolimits_{\ell=1}^{KN}{r_{\ell}}$ w.r.t $\omega_{\ell}$. Therefore, by using Lemma 8 and Lemma 9, we shall design the
Algorithm \ref{Alg:omega} to find the optimal $\bm{\omega}^*$. The basic idea of Algorithm \ref{Alg:omega} is that a larger weight should be assigned to the E2E subchannel with higher increasing rate of achievable rate, because larger weight implies higher power efficiency. The accuracy of Algorithm \ref{Alg:omega} relies on the step size $\triangle{\omega}$.  The smaller $\triangle{\omega}$ is, the more accurate the obtained results are, but the convergence time may become longer. We set $\triangle{\omega}$ as 0.001 in the following discussion.

\begin{algorithm}[h]
\caption{Finding the optimal $\bm{\omega}^*$}
\label{Alg:omega}
\begin{algorithmic}[1]
\FOR{each $\ell \in [1,KN]$}
\STATE Initialize $\omega_{\ell}=0$;\
\ENDFOR
\WHILE {$\sum \omega_{\ell}<1$}
\STATE Find $q=\textrm{arg}\max\limits_{\ell}{\frac{\partial r_{\ell}}{\partial \omega_{\ell}}}$ for all $\ell\in\{1,2,...,KN\}$ in terms of (\ref{Eq:Diffr});\
\STATE Update $\omega_{q}=\omega_{q}+\triangle{\omega}$;\
\ENDWHILE
\STATE Return $\bm{\omega}$.
\label{code:recentEnd}
\end{algorithmic}
\end{algorithm}

\section{Numerical Results}\label{Sec:SectVI}
\begin{figure}
\centering
\includegraphics[width=0.38\textwidth]{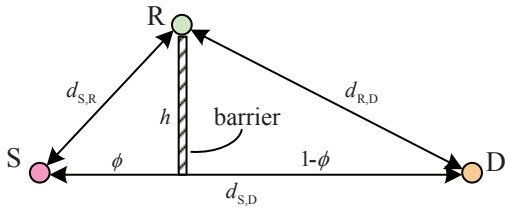}
\caption{Relay position model of two-hop MIMO-OFDM relay network with a barrier between the source and its destination.}\label{Fig:Position}
\end{figure}

In this section,  some numerical results are presented to validate our analysis and discuss the performance of the proposed TSR and PSR. In the simulations, we consider a typical three node relaying network as shown in Figure \ref{Fig:Position}, in which there is a barrier between ${\rm S}$ and ${\rm D}$. ${\rm R}$ is located on the top of the barrier to assist the information delivering from  ${\rm S}$ to ${\rm D}$.

The distance between ${\rm S}$ and ${\rm D}$ is regarded as a reference distance, which is denoted as $d_{\textrm{S},\textrm{D}}$. The height of the barrier is $h$. The variable $\phi\in (0,1)$ is used to describe the ratio of the distance between ${\rm S}$ and the barrier to  $d_{\textrm{S},\textrm{D}}$. In this case, the distance between ${\rm S}$ and ${\rm R}$ and the distance  between ${\rm S}$ and ${\rm R}$ can be respectively expressed by $d_{\textrm{S},\textrm{R}}=\sqrt{h^2+(\phi d_{\textrm{S},\textrm{D}})^2}$ and $d_{\textrm{R},\textrm{D}}=\sqrt{h^2+((1-\phi)d_{\textrm{S},\textrm{D}})^2}$. Note that when $h=0$, $\rm R$ is located on the direct line between $\rm S$ and $\rm D$, which has been widely adopted as a model to discuss the relay position for two-hop relay systems, where the relay is moving along the $\rm{S}-\rm{D}$ direct line.  The total system bandwidth is assumed to be $\mathfrak{B}=5$MHz, so each subcarrier is allocated with $\frac{5\times10^6}{K}$Hz. The power spectral density of the receiving noise at both ${\rm R}$ and ${\rm D}$ is set as $-100$dBm. Besides, the path loss effect is also considered, where the path loss factor is set to 2. The distance between ${\rm S}$ and ${\rm D}$ is regarded as a reference distance, which is set to be 100m. Note that all configuration parameters  mentioned above will not change in the following simulations unless specified otherwise.

To show the performance gain of the optimized PSR and TSR, we consider two schemes with simple configurations as benchmarks, i.e., the simple TSR and the simple PSR. In the simple TSR, the optimal sub-channel pairing is involved, but $\alpha$ is set to a constant with $\alpha=\frac{1}{3}$. $\omega_{\ell}$ and $\varpi_{\ell}$ is assigned by using such a simple strategy, in which the value of each element of $\bm{\omega}$ and $\bm{\varpi}$ is proportional to the eigenvalue of  corresponding sub-channel, i.e., $\omega_{\ell}=\frac{\lambda_{\textrm{S},\ell}}{\sum\nolimits_{j=1}^{KN}{\lambda_{\textrm{S},j}}}$ and $\varpi_{\ell'}=\frac{\lambda_{\textrm{R},\ell'}}{\sum\nolimits_{j=1}^{KN}{\lambda_{\textrm{R},j}}}$.
In the simple PSR, the optimal sub-channel paring is also involved and $\omega_{\ell}$ is determined with a simple method similar to that used in the simple TSR, i.e., $\omega_{\ell}=\frac{\lambda_{\textrm{S},\ell}}{\sum\nolimits_{j=1}^{KN}{\lambda_{\textrm{S},j}}}$.
Moreover, for the optimized TSR, we also consider two different methods for it, i.e., the optimized TSR-I and the  optimized TSR-II. The \textit{optimized TSR-I} denotes the TSR scheme optimized with the alternative updating of $\bm{\omega}$ and $\bm{\varpi}$ described in Lemma 4 and the \textit{optimized TSR-II} denotes the TSR scheme optimized with the approximating optimal $\bm{\omega}^\star$ and $\bm{\varpi}^\star$ described in Lemma 5.

\subsection{Performance vs $\mathcal{P}_{\textrm{S}}$}
\begin{figure}
\centering
\includegraphics[width=0.45\textwidth]{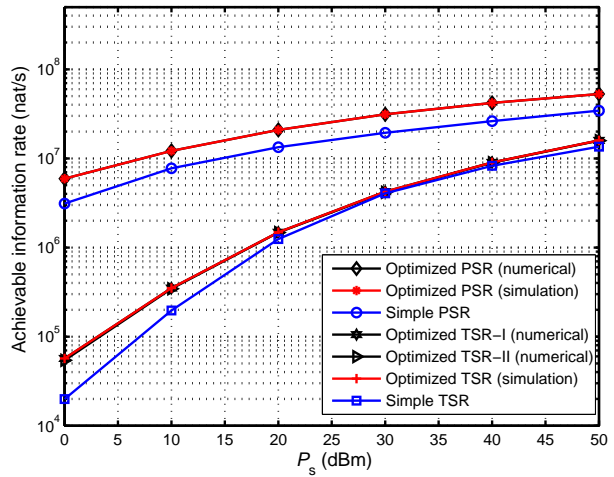}
\caption{System performance vs $\mathcal{P}_{\textrm{S}}$ with $N=2, K=4$ configuration.}\label{Fig:Power2x2}
\end{figure}

\begin{figure}
\centering
\includegraphics[width=0.45\textwidth]{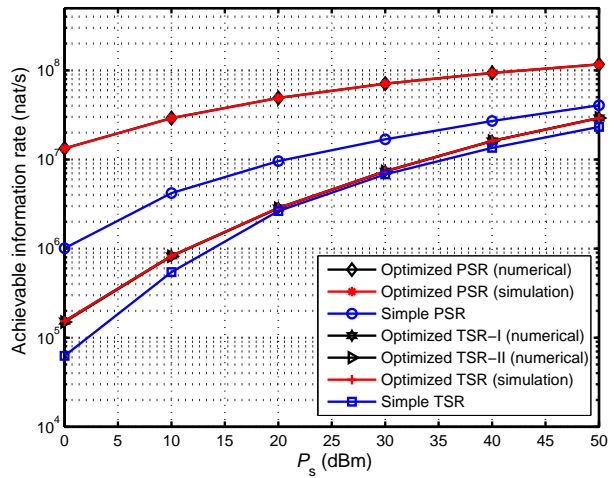}
\caption{System performance vs $\mathcal{P}_{\textrm{S}}$ with $N=4, K=64$ configuration.}\label{Fig:Power4x4}
\end{figure}

In Figure \ref{Fig:Power2x2} and Figure \ref{Fig:Power4x4}, we present the performance of various schemes versus the available transmission power $\mathcal{P}_{\textrm{S}}$ for $N=2, K=4$ and $N=4, K=64$, respectively. In the simulations, $h$ and $\phi$ are set to 0 and 0.3, respectively,  which means that the relay is located on the straight line between ${\rm S}$ and ${\rm D}$ and ${\rm R}$ is closer to ${\rm S}$ than ${\rm D}$. To validate our theoretical analysis, the simulation results obtained by using computer search are also plotted. $\mathcal{P}_{\textrm{S}}$ is changed from 0dBm to 50dBm.

From the two figures, firstly, it can be seen that the numerical results match the simulation ones very well, which indicates the validation of our theoretical analysis and the proposed algorithms. Moreover, it shows that all schemes achieve higher achievable information rate with the increment of $\mathcal{P}_{\textrm{S}}$, because high $\mathcal{P}_{\textrm{S}}$ will lead to high SNR for the system. It is also shown that the performance of the optimized TSR is lower than the optimized PSR. The reason may be explained as follows. Under the same channel conditions, the performances of both TSR and PSR depend on the energy harvested and information received at the relay. In order to well match the harvested energy and collected information, in TSR, besides the power allocation, the time duration for energy transfer and information transmission over all subchannels is adjusted by the same factor $\alpha$. But in PSR, besides the power allocation, each E2E subchannel has its own factor $\rho_i$ ($i=1,..., K$) to adjust the ratio between the energy harvesting and information collecting. Compared with TSR, PSR provides more flexibility to adjust the system resources, so it may yield higher performance gain than TSR.

\begin{figure}
\centering
\includegraphics[width=0.45\textwidth]{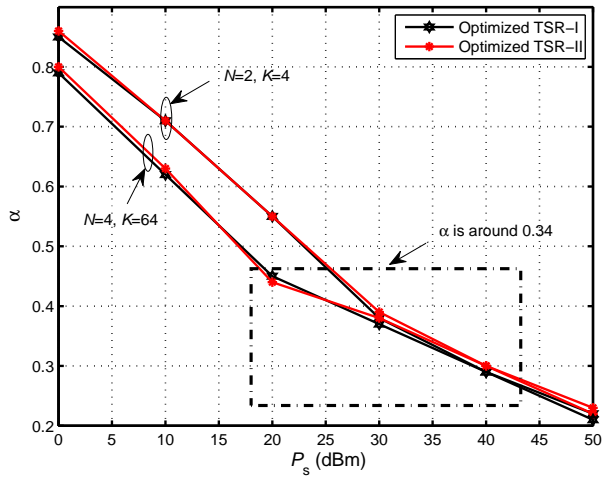}
\caption{The optimal $\alpha^*$ of the optimized TSR vs $\mathcal{P}_{\textrm{S}}$.}\label{Fig:alpha}
\end{figure}

In Figure \ref{Fig:Power2x2}, it also can be observed that when $\mathcal{P}_{\textrm{S}}$ is within the interval of 30dBm to 40dBm, the performance of the simple TSP is very close to that of the optimized ones. The similar results also can be seen in Figure \ref{Fig:Power4x4} between $\mathcal{P}_{\textrm{S}}=$20dBm to $\mathcal{P}_{\textrm{S}}=$30dBm.  The reason can be explained by the results plotted in Figure \ref{Fig:alpha}, where the optimal $\alpha^*$ of TSR is plotted versus $\mathcal{P}_{\textrm{S}}$ for both $N=2, K=4$ and $N=4, K=64$. It can be seen that for $K=N=2$ when $\mathcal{P}_{\textrm{S}}$ is within the interval of 30dBm to 40dBm, the value of the optimal $\alpha^*$ is around 0.34 and for $N=2, K=4$ when $\mathcal{P}_{\textrm{S}}$ is within the interval of 20dBm to 30dBm, the value of the optimal $\alpha^*$ is also around 0.34. Since in the simple TSR, we set $\alpha$ to $\frac{1}{3}$, it is very similar to 0.34, which approximates the optimal ones. Therefore, it makes the performance of the simple TSR very close to that of the optimized one between $\mathcal{P}_{\textrm{S}}=$30dBm to $\mathcal{P}_{\textrm{S}}=$40dBm and between $\mathcal{P}_{\textrm{S}}=$20dBm to $\mathcal{P}_{\textrm{S}}=$30dBm for $K=N=2$ and $K=N=4$, respectively. The intervals of 30dBm to 40dBm  and 20dBm to 30dBm can also be regarded as the efficiently workable intervals for the simple TSR for $N=2, K=4$ and $N=4, K=64$ systems, respectively.

\subsection{Performance vs $\phi$}
\begin{figure}
\centering
\includegraphics[width=0.45\textwidth]{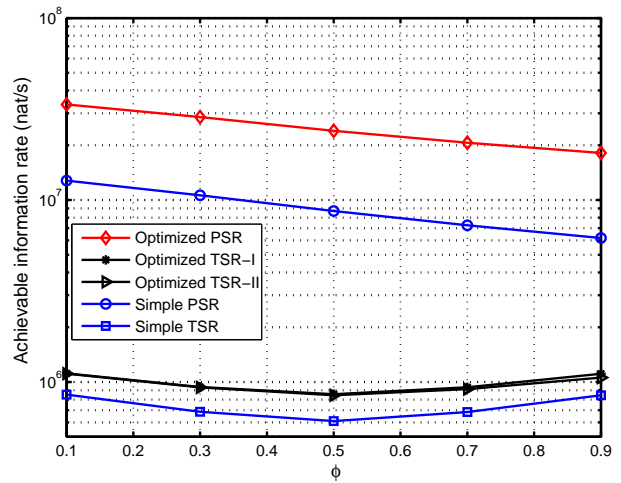}
\caption{System performance vs $\phi$ with $h=25m$, N=3 and N=16.}\label{Fig:position1}
\end{figure}

In this subsection, we shall discuss the performance of our optimized schemes versus the relay location. As illustrated in Figure \ref{Fig:Position}, the relay location is described with the factor $\phi\in (0,1)$, in the simulations of Figure \ref{Fig:position1}, $\phi$ is varied from 0.1 to 0.9.  $h$ is set to $25m$. $K=2$ and $N=2$. $\mathcal{P}_{\textrm{S}}$ is 20dBm. From Figure \ref{Fig:position1}, it can be observed that when the relay moves way from the source, the achievable information rate is decreased. The reason is that the farther the distance between the source and the relay, the less the energy harvesting efficiency at the relay due to the path loss effect. As a result, a relatively lower performance of the system can be achieved when ${\rm R}$ is relatively farther away from ${\rm S}$.

%\begin{figure}
%\centering
%\includegraphics[width=0.45\textwidth]{position-1.eps}
%\caption{System performance vs $\phi$ with \hl{\textit{$h=25m$, N=2 and K=2.}}}\label{Fig:position1}
%\end{figure}

\begin{figure}
\centering
\includegraphics[width=0.45\textwidth]{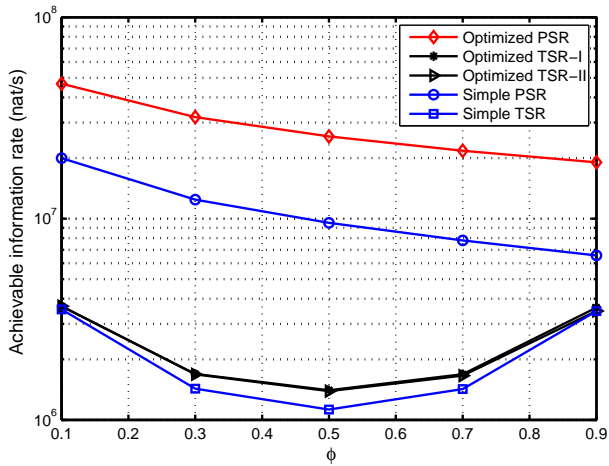}
\caption{System performance vs $\phi$ with $h=0m$, N=3 and K=16.}\label{Fig:position2}
\end{figure}

In the simulations of Figure \ref{Fig:position2}, we set $h=0$. In this case, ${\rm R}$ moves on the straight line between ${\rm S}$ and ${\rm D}$. It can be observed the achievable information rate of the PSR schemes firstly decrease with the growth of $\phi$, while the achievable information rate of the TSR schemes firstly decreases and then increases with the growth of $\phi$. This is the first time to observe such a phenomenon. The reason may be that when $\phi$ is small, ${\rm R}$ is closer to ${\rm S}$, which yields a relatively higher energy harvesting efficiency. When $\phi$ is relatively large, ${\rm R}$ is closer to ${\rm D}$. In this case, although a relatively lower energy harvesting efficiency can be achieved, a relatively better channel quality over the $\rm{R}-\rm{D}$ link is brought, which may improve the system performance.

%\begin{figure}
%\centering
%\includegraphics[width=0.45\textwidth]{position-2.eps}
%\caption{System performance vs $\phi$ with \hl{\textit{$h=0m$, N=2 and K=2.}}}\label{Fig:position2}
%\end{figure}

\begin{figure}
\centering
\includegraphics[width=0.45\textwidth]{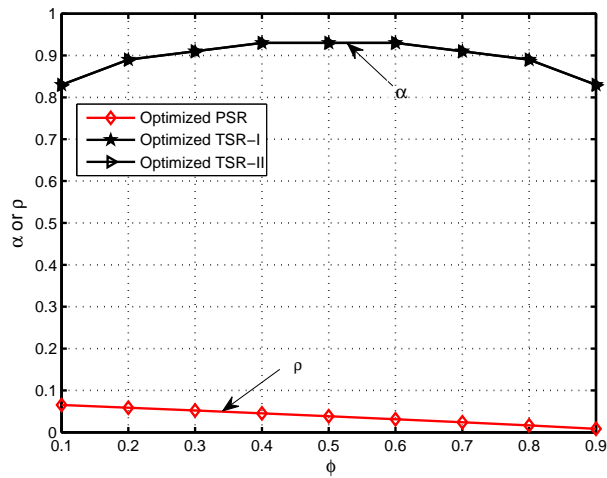}
\caption{System performance vs $\phi$ with $h=0m$.}\label{Fig:ap}
\end{figure}

We also plot the optimal $\alpha$ and $\rho$ in  Figure \ref{Fig:ap} for the case when $K=N=1$ and $h=0$. One can see that with the increment of $\phi$, the value of optimal $\rho$ monotonically decreases while that of $\alpha$ first increases and then decreases.

\subsection{Performance vs the number of antennas $N$}
\begin{figure}
\centering
\includegraphics[width=0.45\textwidth]{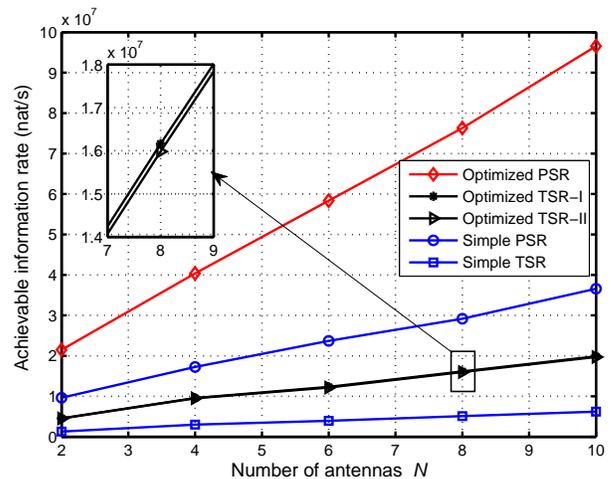}
\caption{System performance vs the number of antennas $N$ with $K=4$ subcarriers.}\label{Fig:AntennaN}
\end{figure}

In this subsection, we shall discuss the impact of the number of antennas $N$ on the system performance. In the simulations, $K$ is set to 4 and $h=0$. We increase $N$ from 2 to 10. Figure \ref{Fig:AntennaN} plots the results averaged over 100 simulations. It can be observed that the achievable information rates of all schemes increase with the increment of the number of antennas. The reason is that more antennas can yield more spatial subchannels. As a result, higher multiplex gain over the subchannels can be achieved.  Moreover, it also shows that PSR achieves the highest achievable information among all schemes and it is also with the highest increasing rate than other ones.

\subsection{Performance vs the number of subcarriers $K$}
\begin{figure}
\centering
\includegraphics[width=0.45\textwidth]{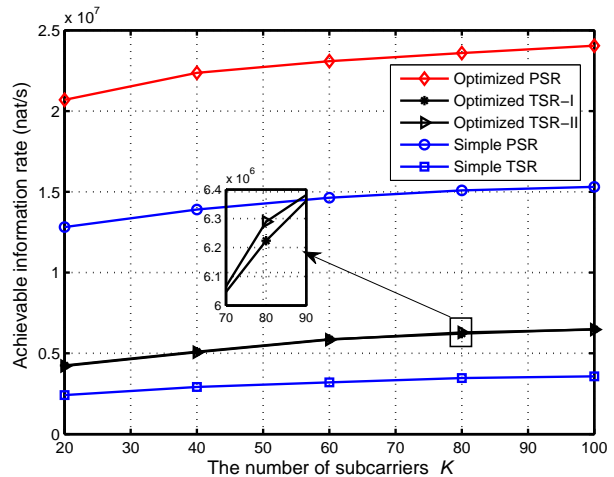}
\caption{System performance vs the number of subcarriers $K$ with $N=2$ antenna configuration.}\label{Fig:SubcarrierK}
\end{figure}

In this subsection, we shall discuss the impact of the number of subcarriers $K$ on the system performance. In the simulations, $N$ is set to 2 and $h=0$. $K$ is gradually increased from 20 to 100. Figure \ref{Fig:SubcarrierK} plots the averaged results over 100 simulations. It can be observed that the achievable information rates of all schemes increase with the increment of the number of subcarrier. However, the increasing rate of each curve goes slower and slower with the increment of $K$. The reason is that more subcarriers may yield more subchannels and bring more flexible configuration to increase the system capacity, but with a fixed total system bandwidth, more subcarrier may cause a smaller bandwidth allocated to each subcarrier. In this case, there exists a trade-off between the number of subcarriers $K$ and system achievable information rate.

% \begin{figure}
%\centering
%\subfigure[$\eta=0.5$]{
%\label{fig:G-STWC:a} %% label for first subfigure
%\includegraphics[width=0.4\columnwidth]{G-1.0.eps}}
%\hspace{1in}
%\subfigure[$\eta=0.5$]{
%\label{fig:G-STWC:b} %% label for second subfigure
%\includegraphics[width=0.4\columnwidth]{G-0.5.eps}}
% \caption{Comparasion of TS-TWR and PS-TWR in terms of $G_{normalized}$.}
%\label{fig:subfig} %% label for entire figure
%\end{figure}

\section{Conclusion}\label{Sec:SectVI}
This paper studied the simultaneous wireless energy harvesting and information transfer for the non-regenerative MIMO-OFDM relaying system, where both the energy harvesting and energy consumption were considered in a single system. We presented two protocols, TSR and  PSR for the system. In order to investigate the system performance limits, we formulated two optimization problems for them to jointly optimize the multiple system configuration parameters so that the end-to-end achievable information rate of each protocol can be maximized. To  the optimized problems, we derived some explicit theoretical results and designed some effective algorithms for them. Various numerical results were presented to confirm our analytical results and to show the performance gain of our optimized PSR and TSR. In addition, it is also shown that the performances of both protocols are greatly affected by the relay position. The achievable information rate of PSR monotonically decreases  with the increment of source-relay distance  and that of TSR firstly decreases and then increases with the increment of source-relay distance and the relatively worse performance is obtained when the relay is placed in the middle of the source and the destination. The simulation results also show that PSR always outperforms TSR in the two-hop non-regenerative MIMO-OFDM system. In addition, increasing either the number of antennas or the number of subcarriers can bring system performance gain to the two protocols.

This work also suggests that in a non-regenerative MIMO-OFDM system, it is better to adopt PSR if the CSI is perfect. For the imperfect CSI case, we will consider it in the future.

%\section*{Acknowledgment}

\begin{appendices}
\section{The Proof of Lemma 1}\label{App:L1}
Applying the eigenvalue decomposition to $\bm{X}_{i}$, we have that $\bm{X}_{i}=\bm{V}_x^{(i)}\bm{\Xi}_x^{(i)}\bm{V}_x^{(i)H}$, where $\bm{V}_x^{(i)}\bm{V}_x^{(i)H}=\bm{\textrm{I}}$ and $\bm{\Xi}_x^{(i)}=\textrm{diag}\Big\{\xi_{x,1}^{(i)},\xi_{x,2}^{(i)},\cdot\cdot\cdot,\xi_{x,N_E}^{(i)}\Big\}$
with $\xi_{x,1}^{(i)}\geq \xi_{x,2}^{(i)}\geq\cdot\cdot\cdot\geq \xi_{x,\textrm{Rank}(\bm{H}_{\textrm{S},i})}^{(i)}\geq0$ and $\sum\nolimits_{q=1}^{N_E}{\xi_{x,q}^{(i)}}\leq \textrm{tr}(\bm{X}_{i})$, where $N_E=\textrm{Rank}(\bm{H}_{\textrm{S},i})$.
Let $\bm{V}_x^{(i)}=[\bm{v}_{x,1}^{(i)},\bm{v}_{x,2}^{(i)},...,\bm{v}_{x,N_E}^{(i)}]^T$ and
$\tilde{\bm{H}}_{\textrm{S},i}=\bm{H}_{\textrm{S},i}\bm{V}_x^{(i)}=[\tilde{\bm{h}}^{(i)}_{\textrm{S},1}, \tilde{\bm{h}}^{(i)}_{\textrm{S},2},\cdot\cdot\cdot,\tilde{\bm{h}}^{(i)}_{\textrm{S},N_E}]$.
It can be deduced that
\begin{flalign}\label{Eq:Tr}
\parallel\bm{H}_{\textrm{S},i}\bm{x}_i\parallel^2&=\textrm{tr}(\bm{H}_{\textrm{S},i}\bm{X}_{i}\bm{H}^H_{\textrm{S},i})=
\textrm{tr}(\tilde{\bm{H}}_{\textrm{S},i}\bm{\Xi}_x^{(i)}\tilde{\bm{H}}^H_{\textrm{S},i})\\
&=\sum\nolimits_{q=1}^{N_E}{\xi_{x,q}^{(i)}}\parallel\tilde{\bm{h}}^{(i)}_{\textrm{S},q}\parallel^2\leq \textrm{tr}(\bm{X}_{i})\parallel\tilde{\bm{h}}^{(i)}_{\textrm{S},1}\parallel^2,\nonumber
\end{flalign}
where the equality holds if
$\parallel\tilde{\bm{h}}^{(i)}_{\textrm{S},1}\parallel^2=\max\limits_{q}\parallel\tilde{\bm{h}}^{(i)}_{\textrm{S},q}\parallel^2$
and
\begin{flalign}
\xi_{x,q}^{(i)}=\left\{ \begin{aligned}
 &\textrm{tr}(\bm{X}_{i}),\,\,q=1,\\
 &0,\quad\quad\,\,\, q=2,3,...,N_E.
 \end{aligned}\right.\nonumber
\end{flalign}
Moreover, since $\bm{H}_{\textrm{S},i}=\bm{U}_{\textrm{S},i}\bm{\Lambda}_{\textrm{S},i}\bm{V}_{\textrm{S},i}^H$,
it can be inferred that $\bm{V}_x^{(i)}=\bm{V}_{\textrm{S},i}$ and $\parallel\tilde{\bm{h}}^{(i)}_{\textrm{S},1}\parallel^2=\max\nolimits_{q}\parallel\tilde{\bm{h}}^{(i)}_{\textrm{S},q}\parallel^2$ only if $\bm{v}_{x,1}^{(i)}$ is the first column of $\bm{V}_{\textrm{S},i}$, which is corresponding to the
largest singular value of $\bm{H}_{\textrm{S},i}$, i.e., $\lambda^{(i)}_{\textrm{\tiny S},1}$. So, $\bm{X}_{i}^{\sharp}=\bm{v}_{s,1}^{(i)}\xi_{x,1}^{(i)}\bm{v}_{s,1}^{(i)H}=\textrm{tr}(\bm{X}_{i})\bm{v}_{s,1}^{(i)}\bm{v}_{s,1}^{(i)H}$.
Lemma 1 is therefore proved.

\section{The Proof of Lemma 2}\label{App:L2}
By applying (\ref{Eq:Tr}) to Problem (\ref{Opt:ProblemX}), Problem (\ref{Opt:ProblemX}) can be equivalently transformed into
\begin{flalign}\label{Opt:tr}
\mathop{\max}\limits_{\textrm{tr}(\bm{X}_1),\textrm{tr}(\bm{X}_2),...,\textrm{tr}(\bm{X}_{K})} &{\sum\nolimits_{i=1}^{K} \textrm{tr}(\bm{X}_{i})\parallel\tilde{\bm{h}}^{(i)}_{\textrm{S},1}\parallel^2}\\
&\sum\nolimits_{i=1}^{K}{\textrm{tr}(\bm{X}_{i})}\leq \mathcal{P}_{\textrm{S}},\,\,\textrm{tr}(\bm{X}_{i})\geq 0.\nonumber
\end{flalign}
By doing so, it can be easily derived that the optimal solution of Problem (\ref{Opt:tr}) is
\begin{flalign}\label{Eq:OpttrX}
\textrm{tr}(\bm{X}_{i})^*=\left\{ \begin{aligned}
 &\mathcal{P}_{\textrm{S}},\,\,i=\textrm{arg} \max\limits_{b=1,...,K}{\parallel\tilde{\bm{h}}^{(b)}_{\textrm{S},1}\parallel^2},\\
 &0,\quad\textrm{otherwise}.
 \end{aligned}\right.
\end{flalign}
Lemma 2 thus is proved.
\section{The Proof of Theorem 2}\label{App:T2}
Let $\mathcal{F}=\sum\nolimits_{\ell=1}^{KN}{r_{\ell}}$, where
\begin{flalign}\label{Eq:rell}
r_{\ell}=\log\bigg(1+\tfrac{(1-\rho_{\ell})\rho_{\ell}\mathcal{P}_{\textrm{S}}^2\omega_{\ell}^2\frac{\lambda_{\textrm{\tiny S},\ell}^2}{\sigma_{\textrm{\tiny R}}^2}\frac{\lambda_{\textrm{\tiny R},\ell'}}{\sigma_{\textrm{\tiny D}}^2}}
{1+(1-\rho_{\ell})\mathcal{P}_{\textrm{S}}\omega_{\ell}\tfrac{\lambda_{\textrm{\tiny S},\ell}}{\sigma_{\textrm{\tiny R}}^2}+\rho_{\ell}\lambda_{\textrm{\tiny S},\ell}{\mathcal {P}}_{\textrm{S}}\omega_{\ell}\tfrac{\lambda_{\textrm{\tiny R},\ell'}}{\sigma_{\textrm{\tiny D}}^2}}\bigg).
\end{flalign}
It can be calculated  that
\begin{flalign}\label{Eq:Rpa1}
&\tfrac{\partial \mathcal{F}}{\partial \rho_{\ell}}=\tfrac{\partial r_{\ell}}{\partial \rho_{\ell}}\nonumber\\
&=\tfrac{A_{\ell}\omega_{\ell}Q_{\ell}\omega_{\ell}(A_{\ell}\omega_{\ell} - 2\rho_{\ell} - 2A_{\ell}\omega_{\ell}\rho_{\ell} + A_{\ell}\omega_{\ell}{\rho_{\ell}}^2 - Q_{\ell}\omega_{\ell}{\rho_{\ell}}^2 + 1)}{(Q_{\ell}\omega_{\ell}\rho_{\ell} + 1)(A_{\ell}\omega_{\ell}( 1- \rho_{\ell}) + 1)(A_{\ell}\omega_{\ell}(1 - \rho_{\ell}) + Q_{\ell}\omega_{\ell}\rho_{\ell} + 1)}.
\end{flalign}
Since $0\leq \rho_{\ell} \leq 1$, it can be easily observed that the denominator of (\ref{Eq:Rpa1}) is always larger than 0. Assume $\frac{\partial \mathcal{F}}{\partial \rho_{\ell}}=0$, for the case $A_{\ell} \neq Q_{\ell}$, we obtain that
\begin{equation}\label{Eq:rho}
\rho_{\ell}=\tfrac{A_{\ell}\omega_{\ell} + 1 - \sqrt{(A_{\ell}\omega_{\ell} +1)( Q_{\ell}\omega_{\ell} + 1)}}{A_{\ell}\omega_{\ell} - Q_{\ell}\omega_{\ell}}.
\end{equation}
It can be easily seen that (\ref{Eq:rho}) satisfies that $0\leq \rho_{\ell} \leq 1$. Moreover, it also can be seen that if $0<\rho_{\ell}<\frac{A_{\ell}\omega_{\ell} + 1 - \sqrt{(A_{\ell}\omega_{\ell} +1)( Q_{\ell}\omega_{\ell} + 1)}}{A_{\ell}\omega_{\ell} - Q_{\ell}\omega_{\ell}}$, $\frac{\partial \mathcal{F}}{\partial \rho_{\ell}}>0$ and if $\frac{A_{\ell}\omega_{\ell} + 1 - \sqrt{(A_{\ell}\omega_{\ell} +1)( Q_{\ell}\omega_{\ell} + 1)}}{A_{\ell}\omega_{\ell} - Q_{\ell}\omega_{\ell}}<\rho_{\ell}<1$, then $\frac{\partial \mathcal{F}}{\partial \rho_{\ell}}<0$. Besides, as $r_{\ell}(0)=r_{\ell}(1)=0$, then we can deduce that $\mathcal {F}$ only has one maximum value which can be achieved only when $\rho_{\ell}$ meets the equation (\ref{Eq:rho}). For the case that $A_{\ell} \neq Q_{\ell}$, (\ref{Eq:rell}) can be simplified as
\begin{flalign}\label{Eq:rell2}
r_{\ell}&=\log\big(1+\tfrac{(1-\rho_{\ell})\rho_{\ell}A_{\ell}\omega_{\ell}Q_{\ell}\omega_{\ell}}
{1+(1-\rho_{\ell})A_{\ell}\omega_{\ell}+\rho_{\ell}Q_{\ell}\omega_{\ell}}\big)=\log\big(1+\tfrac{(1-\rho_{\ell})\rho_{\ell}A_{\ell}^2\omega_{\ell}^2}
{1+A_{\ell}\omega_{\ell}}\big).\nonumber
\end{flalign}
In this case, it can be easily seen that if and only if $(1-\rho_{\ell})\rho_{\ell}$ achieves the maximum, $r_{\ell}$
will be maximal. Obviously, when $(1-\rho_{\ell})=\rho_{\ell}$, i.e., $\rho_{\ell}=0.5$, $(1-\rho_{\ell})\rho_{\ell}$ achieves its maximum value. Therefore, Theorem 2 is proved.
\end{appendices}

\ifCLASSOPTIONcaptionsoff
  \newpage
\fi

\end{document}